\documentclass[11pt]{article}

\usepackage[top=1in, left=1in, right=1in,
bottom=1in]{geometry}

\usepackage{latexsym}
\usepackage[parfill]{parskip}
\usepackage{amssymb,amsmath, bm}
\usepackage{graphicx}
\usepackage{marvosym}
\usepackage{lscape}
\usepackage{rotating}
\usepackage{setspace}
\usepackage{threeparttable}
\usepackage{booktabs}
\usepackage[compact]{titlesec}

\usepackage{caption}
\usepackage{subcaption}

\usepackage{lmodern}
\fontfamily{lmtt}\selectfont
\usepackage[T1]{fontenc}

\titleformat{\section}
  {\normalfont\fontsize{11}{15}\bfseries}{\thesection}{1em}{}

\titleformat{\subsection}
  {\normalfont\fontsize{11}{15}\bfseries}{\thesubsection}{1em}{}

\usepackage{natbib}
\usepackage{color}
\usepackage[pdftex, bookmarksopen=true, bookmarksnumbered=true,
pdfstartview=FitH, breaklinks=true, urlbordercolor={0 1 0}, citebordercolor={0 0 1}]{hyperref}

\usepackage{dcolumn}
\newcolumntype{.}{D{.}{.}{-1}}
\newcolumntype{d}[1]{D{.}{.}{#1}}

\usepackage{amsthm}

\newtheorem{assumption}{Assumption}

\newcommand{\R}{\ensuremath{\mathbb{R}}}

\newcommand{\E}{\ensuremath{\mathbb{E}}}

\newcommand{\Var}{\text{Var}}
\newcommand\var[1]{\Var\left\{ #1 \right\}}

\newcommand{\logit}{\text{logit}}

\def\b1{\boldsymbol{1}}

\begin{document}

\pagestyle{plain}

\newcommand{\blind}{0}

\newcommand{\tit}{\Large Hospital Quality Risk Standardization via Approximate Balancing Weights}

\if0\blind

{\title{\tit\thanks{
We thank Peng Ding, Skip Hirshberg, and Sam Pimentel for helpful feedback, as well as seminar participants in the Penn Causal Group.
The dataset used for this study was purchased with a grant from the Society of American Gastrointestinal and Endoscopic Surgeons. Although the AMA Physician Masterfile data is the source of the raw physician data, the tables and tabulations were prepared by the authors and do not reflect the work of the AMA. The Pennsylvania Health Cost Containment Council (PHC4) is an independent state agency responsible for addressing the problems of escalating health costs, ensuring the quality of health care, and increasing access to health care for all citizens. While PHC4 has provided data for this study, PHC4 specifically disclaims responsibility for any analyses, interpretations or conclusions. Some of the data used to produce this publication was purchased from or provided by the New York State Department of Health (NYSDOH) Statewide Planning and Research Cooperative System (SPARCS). However, the conclusions derived, and views expressed herein are those of the author(s) and do not reflect the conclusions or views of NYSDOH. NYSDOH, its employees, officers, and agents make no representation, warranty or guarantee as to the accuracy, completeness, currency, or suitability of the information provided here. This publication was derived, in part, from a limited data set supplied by the Florida Agency for Health Care Administration (AHCA) which specifically disclaims responsibility for any analysis, interpretations, or conclusions that may be created as a result of the limited data set. The authors declare no conflicts. Eli Ben-Michael and Avi Feller gratefully are funded by National Science Foundation Grant \#1745640. Rachel Kelz is funded by a grant from the National Institute on Aging, R01AG049757- 01A1.}}
\author{Luke Keele\thanks{University of Pennsylvania, Philadelphia, PA, Email: luke.keele@gmail.com}
\and Eli Ben-Michael\thanks{University of California, Berkeley, Berkeley, CA, Email: ebenmichael@berkeley.edu}
\and Avi Feller\thanks{University of California, Berkeley, Berkeley, CA, Email: afeller@berkeley.edu}
\and Rachel Kelz\thanks{University of Pennsylvania, Philadelphia, PA, Email: Rachel.Kelz@pennmedicine.upenn.edu}
\and Luke Miratrix\thanks{Harvard University, Cambridge, MA, Email: lmiratrix@g.harvard.edu}
}

\date{\today}

\maketitle
}\fi

\if1\blind
\title{\bf \tit}
\maketitle
\fi

\begin{abstract}
Comparing outcomes across hospitals, often to identify underperforming hospitals, is a critical task in health services research. However, naive comparisons of average outcomes, such as surgery complication rates, can be misleading because hospital case mixes differ --- a hospital's overall complication rate may be lower due to more effective treatments or simply because the hospital serves a healthier population overall. In this paper, we develop a method of ``direct standardization'' where we re-weight each hospital patient population to be representative of the overall population and then compare the weighted averages across hospitals. Adapting methods from survey sampling and causal inference, we find weights that directly control for imbalance between the hospital patient mix and the target population, even across many patient attributes.  Critically, these balancing weights can also be tuned to preserve sample size for more precise estimates. We also derive principled measures of statistical precision, and use outcome modeling and Bayesian shrinkage to increase precision and account for variation in hospital size. We demonstrate these methods using claims data from Pennsylvania, Florida, and New York, estimating standardized hospital complication rates for general surgery patients. We conclude with a discussion of how to detect low performing hospitals.
\end{abstract}

\begin{center}
\end{center}

\clearpage
\doublespacing


\section{Introduction: Judging Hospital Quality}

How can we assess quality across hospitals? Simple comparisons of hospital-specific outcomes can be misleading: Hospitals that treat patient populations with complex, chronic conditions will generally have worse outcomes than hospitals that treat patients who are healthier prior to surgery. Thus, a hospital's outcomes may be better due to more effective treatments or simply because the hospital serves a healthier clientele. Risk adjustment, also known as risk standardization, refers to a set of statistical methods that adjust the hospital patient mix to make hospital outcomes more comparable \citep{normand2007statistical}. Risk standardization is widely used to evaluate hospitals and provide the public with information on hospital quality. For example, Medicare's online tool, Hospital Compare, uses risk standardization to help patients identify high quality hospitals. Risk standardization comes in two forms often referred to as direct and indirect standardization. 

In this paper, we develop a weighting-based approach to direct standardization. In our approach, we view each hospital's patient population as a non-representative sample from the overall patient population. We then generate a set of weights for each hospital so the weighted distribution of its patients matches the overall population. We show that this form of direct standardization reduces bias due to systematic differences in patient populations across hospitals while maintaining precision.

Our method is inspired in part by ``template matching'' \citep{silber2014template} where an identified set of patients at each hospital are chosen to closely match the set of template patients based on a canonical list of patient characteristics. Compared to template matching, we show substantial gains in both bias control and precision. We also identify a bias-precision tradeoff, and find that by regularizing the weights we can substantially increase precision in our hospital specific estimates while only incurring what appears to be a small increase in bias. We can demonstrate how to use an outcome model both to reduce remaining bias and to improve the precision of the hospital quality estimates. Finally, we apply a Bayesian shrinkage estimator as an additional step in order to better account for variability in the size of hospitals.

We use our method to better understand hospital quality for general surgical procedures using claims data from Pennsylvania, New York, and Florida. Overall, we find that patient mix plays a major role: differences in patient mix explain about 70\% of the variation in unadjusted hospital complication rates. In particular, we identify hospitals that appear to be high quality but that generally serve healthy patients; after adjustment, their estimated complication rate is quite poor. Conversely, we identify hospitals that appear to be low quality but that actually have relatively good complication rates after adjusting for their patient mix.

Our paper proceeds as follows. We first discuss our application in greater detail, and also review methods of direct standardization.
In Section~\ref{sec:weight_main}, we outline our approach of using balancing weights, a tool taken from the literature on causal inference in observational studies, for risk standardization. We then derive methods of variance estimation that are consistent with our estimated weights. 
These sections form the core of our approach.  Next, we apply regression modeling to adjust for remaining imbalance and increase precision in the hospital level estimates. We then outline how to apply a Bayesian shrinkage estimator to account for variation in hospital size and to obtain improved estimates of hospital performance. We apply our methods to the claims data on general surgery. We then conclude.

\subsection{Hospital quality in PA, FL, and NY on General Surgical Performance}
\label{sec:application_intro}

General surgery consists of high volume surgical procedures that are conducted in almost all hospitals, including procedures such as appendectomy (removal of the appendix), cholecystectomy (gall bladder), mastectomy (breast), and hernia repairs. Since deaths are rare in general surgery, we use postoperative complications (e.g., infections and bleeds) as an indication of a problematic surgical procedure. We assess hospital quality in general surgery by estimating the risk-adjusted rates of such complications.

In our analysis, we use data based on all-payer hospital discharge claims in New York, Florida, and Pennsylvania from 2012-2013. The data contain patient sociodemographic and clinical characteristics, including a measure of patient frailty, an indicator for sepsis, and 31 indicators for comorbidities based on Elixhauser indices \citep{elixhauser1998comorbidity}, as well as admission type (emergency, urgent, or elective), type of insurance, and age. We analyze 44 general surgery operations.\footnote{We restrict the patient population to those patients who had a surgical procedure included in the Agency for Healthcare Research and Quality (AHRQ) Clinical Classifications Software (CCS). CCS categories uses International Classification of Diseases, 9th revision, Clinical Modification (ICD-9-CM) diagnosis and procedure codes to classify whether procedures are surgical or not \citep{decker2014specialization}. We also removed any hospitals that performed fewer than 30 procedures over the two-year period, which removed 70 hospitals (out of 593) and 605 patients (out of 622,272).} Across the three states, we have a total of 621,667 patients in 523 hospitals with between 30 and over 8000 general surgery cases for the study period.  The median number of patients was 700. Our primary outcome of interest is a binary indicator for the development of one or more complications after general surgery (identified using ICD-9-CM diagnosis codes). Our goal is to develop risk-standardized measures of quality for these hospitals based on observed complications. Next, we motivate the need for risk adjustment using our data.

\begin{figure}
\centering
\includegraphics[width=0.5\textwidth]{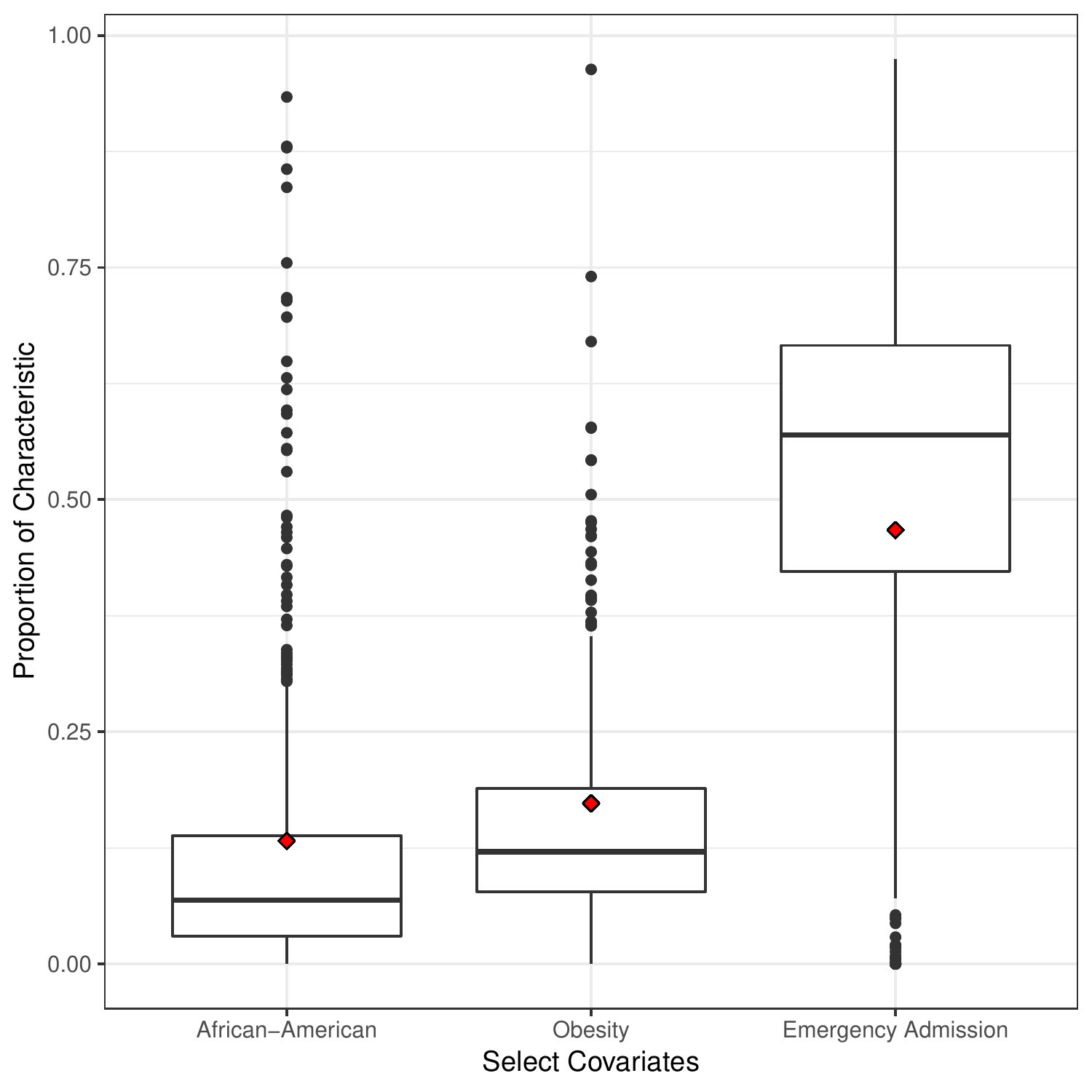}
\caption{Boxplots of Hospital Casemix Distributions. Diamond represents population mean for that covariate.}
\label{fig:1}
\end{figure}

Figure~\ref{fig:1} displays boxplots of hospital-level proportions of three key patient characteristics: whether a patient is African-American, whether a patient is obese--BMI greater than 30, and whether the procedure was an emergency admission. All three characteristics are important predictors of complications in the cohort. As the boxplots show, there is substantial variation in all three attributes. For instance, only 17 percent of patients in the sample are obese, while several hospitals have patient populations in which more than half are obese. The goal of standardization is to adjust for differences in patient mix like these, allowing us to more directly compare outcomes across hospitals.

\section{Direct Risk Standardization for Comparing Hospital Outcomes}

There is a large literature in statistics and health services research on risk adjustment via standardization; see \citet{normand2007statistical,normand2016league} for reviews. There are two main types of risk adjustment: indirect and direct.  At a high level, indirect standardization asks, ``how should this hospital have done, given the patients they serve?'' Direct standardization asks, ``how would this hospital do, if given the same types of patients as everyone else?'' See \citet{longford2019performance} for a discussion of how risk adjustment via both indirect and direct standardization can be viewed as a causal inference problem in the potential outcomes framework. In this paper, we focus on direct standardization, which we review next.

Direct standardization adjusts for the hospital case mix, typically by weighting outcomes to a target distribution. This allows the analyst to target the differences between hospital covariate distributions and the population covariate distributions \citep{george2017mortality}. Direct standardization methods, however, have traditionally been limited by the inability to incorporate more than a few patient level variables \citep{iezzoni2012risk}. For example, in many direct standardization analyses, outcomes are risk adjusted only for age. 

Template matching is a recently developed form of direct standardization that can easily risk adjust for many patient level covariates \citep{silber2014hospital,silber2014template,silber2016auditing}. It makes the goals of hospital standardization clear, and focuses the researcher's attention on important considerations, such as whether the covariates are sufficient to justify an attempt at standardization. Under template matching, the investigator seeks to understand how different hospitals would perform with patients similar to a sample (``template'') of patients. For each hospital, matching methods are used to find a subset of patients that are highly comparable to the template. Hospital quality is then evaluated based on this matched set. One potential way to construct the template is to use patients that are most representative of the overall population. Template matching serves as an important breakthrough in methods for standardization as it both avoids the strong parametric assumptions of models needed for indirect standardization and provides an estimate of hospital quality that does not rely on a ratio estimate. 

Template matching, however, may be inefficient since it only uses a limited part of the available data. Furthermore, when hospital sizes vary widely, the mechanics of template matching become more complex \citep{silber2016auditing}. Finally, extant work on template matching has not addressed best practice for estimating statistical precision. We address these issues via direct standardization using weighting methods.


\section{Direct Standardization via Approximate Balancing Weights}
\label{sec:weight_main}

We now develop a weighting method for direct standardization, which solves a convex optimization problem to simultaneously optimize for balance and effective sample size. First, we review weighting methods. Next, we outline notation and assumptions and then turn to specific implementation details.


\subsection{Review: Balancing weights}
\label{sec:review_bal_weights}

Weighting methods have a long history in survey sampling and causal inference \citep{horvitz1952generalization, lohr2009sampling, robins1995semiparametric}.
The goal is to weight target groups so they have a similar distribution on a given set of covariates.
The reweighted groups can then be directly compared on some outcome of interest, since the reweighted groups are comparable on baseline characteristics.
If we were only interested in balancing a small number of patient characteristics, we could directly apply classical calibration approaches from survey sampling \citep[see e.g.][]{Deming1940, Deville1992, Deville1993}. In this case, the resulting weights would achieve exact balance where the re-weighted and target covariate averages are equal. 

In our setting, however, we need to find weights that balance a large number of patient characteristics and comorbidities, so achieving exact balance in infeasible, especially for smaller hospitals. One alternative is to use a traditional Inverse Probability Weighting (IPW) estimator, which is widely used to estimate treatment effects \citep{Robins:2000, Imbens:2004}.
This approach can be somewhat unstable, however, resulting in re-weighted samples that still are not well balanced. To avoid such issues, we build on recent advances in the causal inference literature where analysts use \emph{approximate balancing} weights that are designed to directly target covariate balance in the estimation process. This class of weighting methods solve a convex optimization problem to find a set of weights that target a specific loss function \citep{Hainmueller2011,Zubizarreta2015}. See \citet{benmichael_balancing_review} for a recent review of these weighting methods.

\subsection{Hospitals as Non-Representative Samples}

In our data, we observe $i = 1, \ldots, n$ patients nested in hospitals $j = 1,\ldots,J$, with patient hospital indicator $Z_i \in \{1,\ldots,J\}$ and $n_j$ patients in each hospital.\footnote{In principle, each observation is a patient-surgery pair. Since we focus only on one surgery per patient, we ignore this complication in our exposition.} For each patient, we observe a vector of background covariates $X_i \in \mathbb{R}^d$. We also observe an outcome $Y_i$, which in our data is a binary indicator for a postoperative complication. The primary statistical problem is that the distribution of patient- and surgery-level characteristics vary across hospitals --- $p(x \mid Z = j) \neq p(x \mid Z = j')$ for $j \neq j'$ --- so the difference between the average outcomes between two hospitals reflects both differences in hospital quality and differences in the distribution of patient attributes.

Formally, we denote the expected value of our outcome given observed covariates $x$ and hospital $j$ as $m_j(x) = \E[Y \mid X = x, Z = j]$. 
We can think of $m_j(x)$ as a ``quality surface'' of the hospital: it describes our expected outcome for hospital $j$ when serving a patient with characteristics $x$.
The expected overall average outcome in hospital $j$ is then 
\[ \rho_j = \E[Y \mid Z = j] = \int m_j(x) dP(x \mid Z = j) . \]
This quantity is easily estimated by the raw mean of hospital $j$, $\bar{Y}_j \equiv \frac{1}{n_j} \sum_{Z_i = j} Y_i$.
However, these estimates are not directly comparable across hospitals. Even if two hospitals $j$ and $j'$ have identical quality surfaces, $m_j(x) = m_{j'}(x)$, $\rho_j$ and $\rho_{j'}$ may differ as they average the quality surfaces over different distributions.

The goal of risk adjustment is to remove this dependence between the patient and surgery characteristics $X$ and the hospital $Z$. We do this by considering a set of hospital estimands that each take the expectation of $m_j(x)$ over a common distribution $X \sim P^*$:
\begin{equation}
  \label{eq:general_mu_j}
  \mu_j = \int m_j(x) dP^*(x).
\end{equation}
These $\mu_j$ are more directly comparable as we have removed systematic differences in distribution. There are many distributions that we may consider shifting towards, e.g., a region of high overlap between hospital patient distributions, or the marginal distribution of patients. Here, we focus on one simple estimand: the empirical distribution of the covariates across all hospitals. This gives
\begin{equation}
  \label{eq:muj}
  \mu_j = \frac{1}{n}\sum_{i = 1}^n m_j(X_i).
\end{equation}
We can view this estimand as the expected outcome of hospital $j$ if its patient mix were the same as the full population of patients. It is important to note that estimands for direct standardization differ from estimands under indirect standardization. Our estimand examines how hospital performance differs from expected across patient characteristics, giving the same weight for each patient type to each hospital. By contrast, indirect standardization examines how each hospital differs from expected, given the model, from those patients that hospital serves.

An important question is how to interpret risk adjusted differences in outcomes across hospitals. One can view risk-adjusted quality measures as being informative of hospital quality without giving them a causal interpretation. With additional assumptions, however, a causal interpretation is possible \citep{longford2019performance}. Specifically, we would need to assume that differences in hospital patient mix are fully captured by $X$, which implies that unobserved differences in patient mix do not contribute to the estimates. This assumption would be violated if the patient mix at some hospitals was significantly at higher risk for complications, but this elevated risk was not captured by the patient level covariates. Understanding the possible role of unobservable differences is critical if risk adjustment is the basis for targeting hospitals for improvement efforts; see \citet{hull2018estimating} for further discussion.

Even if we wish to perform a non-causal comparison of hospitals with different patient populations, we still have to impose the additional assumption that, at least in principle, any type of patient (as defined by $X$) in our reference distribution $P^*$ could receive care at any hospital.
We formalize this as an overlap assumption:\footnote{We could further restrict our estimand to a set of patients where there is overlap by restricting $P^*$. We leave determining a common $P^*$ in the case of partially overlapping distribution to future work.}
\begin{assumption}[Overlap]
  \label{a:overlap}
  $0 < P(Z = j \mid X = x)$ if $P^*( x ) > 0$
\end{assumption}
Assumption \ref{a:overlap} rules out the possibility that a hospital would never treat a particular type of patient.
For example, we assume no hospital treats only women and that all hospitals perform the full range of surgeries we are investigating.
As we discuss in Section \ref{sec:optimization_main}, the distribution of estimated weights is a useful diagnostic for assessing overlap in practice.

\subsection{Estimating hospital means}

With direct standardization, we estimate the average population outcome for hospital $j$, $\mu_j$, with a weighted average of observed outcomes for hospital $j$, using normalized weights $\hat{\gamma}$:
\begin{equation}
  \label{eq:muj_hat}
  \hat{\mu}_j = \sum_{Z_i=j} \hat{\gamma}_i Y_i,
\end{equation}
with $\sum_{Z_i=j} \hat{\gamma} = 1$.

Our setup accommodates general function classes for the quality function $m_j(x)$; see \citet{kallus2016generalized}, \citet{hirshberg2019minimax}, and \citet{Hazlett2019} for further discussion.  To motivate our optimization we impose the simplifying restriction that the quality function $m_j(x)$ is a linear function of some transformation of the covariates: 
\begin{equation}
  \label{eq:quality_curve_linear}
  m_j(x) = \alpha_j + \beta_j \cdot \phi(x),
\end{equation}
with $\phi:\R^d \to \R^p$ and $\beta_j \in \R^p$.
Thus, $\phi(x)$ is our basis and is how we represent the information the covariates provide.
We discuss our choice of basis for our application in Section \ref{sec:results}.

Given $m_j(x)$, $\varepsilon_i \equiv Y_i - \beta_j \cdot \phi(X_i)$ is the \emph{residual} of outcome $Y_i$ given covariates $X_i$ and hospital $Z_i = j$. We can then express the difference between the weighted average in hospital $j$, $\hat{\mu}_j$ and the target estimand $\mu_j$:
\begin{equation}
  \label{eq:weighted_error}
  \hat{\mu}_j - \mu_j = \underbrace{\beta_j \cdot \left(\frac{1}{n_j}\sum_{Z_i = j} \hat{\gamma}_i \phi(X_i) - \bar{\phi} \right)}_{\text{bias}} + \underbrace{\sum_{Z_i = j} \hat{\gamma_i} \varepsilon_i}_{\text{variance}},
\end{equation}
with $\bar{\phi} \equiv \frac{1}{n}\sum_{i=1}^n \phi(X_i)$ the overall population mean of the covariate vector $\phi(x)$.

The error in \eqref{eq:weighted_error} has two components: (1) systematic bias due to imbalance in $\phi(X_i)$ between hospital $j$ and the overall sample; and (2) idiosyncratic error due to noise. The goal is to find weights that control both terms. 

For the first term in Equation \eqref{eq:weighted_error}, the challenge is that the coefficients $\beta_j$ are unknown. Using the Cauchy-Schwarz inequality, we can see that controlling the imbalance in $\phi(X_i)$ also controls the systematic bias (conditional on $X$ and $Z$):

\begin{equation}
  \label{eq:bias}
  \left|\E\left[\hat{\mu}_j - \mu_j\mid X, Z = j\right]\right| \leq \|\beta_j\|_2\underbrace{\left\|\frac{1}{n_j}\sum_{Z_i = j} \hat{\gamma}_i  \phi(X_i) - \bar{\phi}\right\|_2}_{\text{Imbalance}}.
\end{equation}

Thus, under Equation \eqref{eq:weighted_error}, reducing the imbalance controls the bias regardless of the true $\beta_j$.
If there were only a small number of patient characteristics, we could likely achieve exact balance, where the imbalance term in Equation \eqref{eq:bias} is exactly zero. This is not feasible in settings with a richer set of covariates; the central goal of the optimization problem below is to make the imbalance term as small as possible, all else equal.

For the second term in Equation \eqref{eq:weighted_error}, the challenge is that the individual $\varepsilon_i$ are unknown. However, we can bound the variance (conditional on $X$ and $Z$) by the sum of the squared weights:
\begin{equation}
  \label{eq:variance}
  \Var\left(\hat{\mu}_j - \mu_j \mid X, Z = j\right) = \frac{1}{n_j^2}\hat{\gamma}' \Sigma_j \hat{\gamma} \leq \frac{\lambda_j}{n_j^2} \sum_{Z_i = j} \hat{\gamma}_i^2,
\end{equation}
where $\Sigma_j$ is the variance-covariance matrix of the noise terms, and $\lambda_j$ is the maximum eigenvalue of $\Sigma_j$.
This shows that we can reduce the variance by limiting the spread of the estimated weights. 
That is, the more homogenous we can make the weights, the more precise the resulting estimates. Here, we choose to penalize the sum of the squared weights, $\sum_{Z_i=j}\hat{\gamma}_i^2$, though other penalties are possible; see \citet{benmichael_balancing_review}.

\subsection{Weighting via convex optimization}
\label{sec:optimization_main}

We can now combine these two objectives into the following optimization problem:
\begin{equation}
  \label{eq:objective}
  \begin{aligned}
    \min_\gamma \;\; & \sum_{j=1}^J \left[\left\| \bar{\phi} - \sum_{Z_i = j} \gamma_i \phi(X_i) \right\|_2^2 + \lambda n_j \sum_{Z_i = j} \gamma_i^2\right]\\
    \text{subject to  } 
    & \sum_{Z_i = j} \gamma_i = 1\\
    & \ell \leq \gamma_i \leq u 
  \end{aligned} .
\end{equation}

The optimization problem \eqref{eq:objective} trades off two competing terms for each hospital $j$: better balance (and thus lower bias) and more homogeneous weights (and thus lower variance).\footnote{For a single hospital, the objective in optimization problem \eqref{eq:objective} reduces to a special case of the minimax linear estimation proposal from \citet{hirshberg2019minimax}, with a particular choice of function class.
The above extends to the case with multiple hospitals.} A global hyperparameter $\lambda$ negotiates the tradeoff: when $\lambda$ is large the optimization problem will prioritize variance reduction and search for more uniform weights, when $\lambda$ is small it will instead prioritize bias reduction.
We explore the role of $\lambda$ in the bias-variance tradeoff empirically in Section \ref{sec:lambda}.

The constraint set in Equation \eqref{eq:objective} has two components. First, we constrain the weights to sum to one within each hospital, ensuring that each hospital estimate is in fact a weighted average of its outcomes. Second, we constrain the weights to have lower bound $\ell$ and upper bound $u$. We set the lower bound $\ell = 0$ so that weights are non-negative and do not extrapolate outside of the support of the data. Combined with the sum-to-one constraint, each individual weight $\hat{\gamma}_i$ corresponds to the fraction of hospital $Z_i$'s outcome dictated by unit $i$.  These constraints also stabilize the estimate by ensuring sample boundedness; for example, $\hat{\mu}_j$, which estimates a complication rate, is always a valid proportion, $\hat{\mu}_j \in [0,1]$.

Under our weighting approach, extreme weights will signal if the covariates for a specific hospital do not overlap with the patient population. We can target such extreme weights using the upper bound $u$. Setting the upper bound to $u < 1$ would prevent the optimization problem from putting too much weight on any single patient in a hospital. For example, setting $u = 0.2$, would ensure that we do not put more than 20\% of the weight on any individual patient. As such, investigators can evaluate the overlap assumption in the estimation process --- without reference to outcomes --- by inspecting extreme weights and assessing the sensitivity of the estimates to the choice of upper bound $u$. In our primary results, we set $u = 1$ (no constraint) and investigate the impact of setting $u$ to be less than 1 in the supplementary material. 

Finally, the optimization problem of \eqref{eq:objective} obtains $\hat{\gamma}_i$ without using outcome information. Similar to matching and propensity score methods in observational studies, this is a design step where we set up our final evaluation using covariate information alone. We can then simply estimate the adjusted hospital means by taking a weighted average of the patient outcomes, $\hat{\mu}_j = \sum_{Z_i=j} \hat{\gamma}_i Y_i$.

\subsection{Variance estimation}
\label{sec:variance_estimation}

Once we estimate weights, we can quantify uncertainty using standard results from survey sampling. Under the assumption that individual outcomes are independent within a hospital, the sampling variance (conditional on the weights) is:
\[ 
\var{ \hat{\mu}_j | \hat{\gamma}_j } = \var{ \sum_{Z_i = j} \hat{\gamma}_j Y_i } =  \sum_{Z_i = j} \hat{\gamma}_j^2 \var{ Y_i } .
\]

For each hospital, we could then estimate this variance using a plug in:
\begin{align}
 \widehat{\text{se}}( \hat{\mu}_j | \hat{\gamma}_j ) = \left[ \hat{\sigma}^2_j  \sum_{Z_i = j} \hat{\gamma}_j^2 \right]^{1/2} = \frac{\hat{\sigma}_j}{ \sqrt{ n^{\text{eff}}_{j} } } , \label{eq:se}
\end{align}
where we estimate the variance of the outcomes as
\[ \hat{\sigma}_j^2 = \frac{1}{\sum_{Z_i = j} \hat{\gamma}_i^2 - 1 }\sum_{Z_i = j} \hat{\gamma}_i^2 (Y_{ij} - \hat{\mu}_j)^2 \]
and where
\[ n^{\text{eff}}_{j} \;\; \equiv \;\; \left(\sum_{Z_i = j} \hat{\gamma}_i\right)^2 \Bigg/ \sum_{Z_i = j} \hat{\gamma}_i^2 \;\; = \;\; 1 \Bigg/ \sum_{Z_i = j} \hat{\gamma}_i^2 \]
is the \emph{effective sample size} for hospital $j$ \citep{potthoff1992equivalent,lohr2009sampling}.\footnote{The effective sample size is the inverse of the dispersion penalty in the balancing weights optimization problem in Equation \eqref{eq:objective}.}
The last equality above assumes the weights sum to 1 within hospital, which holds under the constraint in the optimization problem in Equation \eqref{eq:objective}.

If all of the hospitals in our sample were large, the individual $\sigma_j$, estimated separately for each hospital, would be stable. In practice, however, the $\hat{\sigma}_j$ estimates from smaller hospitals may be noisy, which will complicate subsequent adjustments, especially partial pooling across hospitals.
In particular, if a small hospital has an unusually small estimated standard error, its point estimate will receive excessive weight when trying to estimate cross-hospital variation. In our empirical example, for instance, some hospitals had na\"ive estimates of 0 for the standard error since they had no complications observed.

Therefore, we instead pool the individual standard deviation estimates into a global estimate.
Specifically, we estimate the pooled standard deviation as
\[ 
\hat{\sigma}^2_{\text{pool}} = \frac{1}{N^{\text{eff}}} \sum_{j=1}^J n^{\text{eff}}_{j} \hat{\sigma}_j^2,
\]
where $N^{\text{eff}} = \sum_j n^{\text{eff}}_j$ is the pooled effective sample size. The pooled variance is a weighted average of the noisy hospital-specific variance estimates. The hospital specific standard errors are then $\hat{\sigma}^2_{\text{pool}}/\sqrt{n^{\text{eff}}_{j}}$. See \citet{Weiss:2017hz} for an extended discussion of this approach for stabilizing estimates of impact variation in multisite trials; they find that the potential bias from ignoring heteroskedasticity is small. If heteroskedasticity were a concern, we could also merge this variance estimation step with a Bayesian model, as discussed below.

\section{Extensions}

We now consider two extensions to the basic weighting approach: bias correction and partial pooling. These can be used together or separately to improve risk adjustment based on weighting alone. 

\subsection{Incorporating additional bias correction}
\label{sec:covariate_adjustment}

As with other methods of direct standardization, the weighted estimator in Section \ref{sec:optimization_main} has the benefit of being \emph{design-based}, that is, the weights $\hat{\gamma}$ solving the optimization problem in Equation \eqref{eq:objective} are independent of the outcomes \citep{rubin2008objective}. There are, however, reasons to utilize outcome information in the risk adjusted estimates. We now describe how to include outcome information into the hospital level estimates.

Especially in smaller hospitals, the weighted mix of patients in hospital $j$ may still not quite match the target distribution. We therefore use a model to estimate how far off our weighted average outcome might be, given this remaining imbalance, and subtract that estimated bias off.
In particular, given an estimate of the quality surface, $\hat{m}_j(x)$, for hospital $j$ we adjust our estimated outcome as follows:
\begin{equation}
  \label{eq:bias_correct}
  \hat{\mu}_j = \sum_{Z_i = j}\hat{\gamma}_i Y_i + \underbrace{\sum_{i=1}^n \hat{m}_j(X_i) - \sum_{Z_i = j} \hat{\gamma}_i\hat{m}_j(X_i)}_{\text{imbalance in } \hat{m}_j(\cdot)}.
\end{equation}
Analogous to bias correction for matching \citep{rubin1973bias,abadie2011bias}, the bias is estimated as the imbalance in the estimated quality surface, $\hat{m}_j(\cdot)$. This bias is then removed from the risk adjusted hospital outcome. If $\hat{m}_j(\cdot)$ is a good estimator for the true quality surface $m_j(\cdot)$, then the adjustment term in Equation \eqref{eq:bias_correct} will reduce any bias due to remaining imbalance \citep[see][for more discussion on bias-corrected balancing weights]{Athey2018a,Hirshberg2019_amle}. We obtain our $m_j(\cdot)$ using least squares, but could instead use more flexible non parametric estimator.

This formulation can also lead to improved precision, since this removes additional variation due to imbalance, and to improved \emph{estimation} of that precision, due to more accurate estimates of residual variation. Borrowing from model-assisted survey sampling \citep{sarndal2003model, Breidt2017}, we can re-write Equation~\ref{eq:bias_correct} as
\begin{equation}
  \label{eq:model_asssist}
  \hat{\mu}_j = \sum_{i=1}^n \hat{m}_j(X_i) +  \sum_{Z_i = j} \hat{\gamma}_i \left(  Y_i  - \hat{m}_j(X_i) \right) = \sum_{i=1}^n \hat{m}_j(X_i) +  \sum_{Z_i = j} \hat{\gamma}_i \hat{\epsilon}_i ,
\end{equation}
where the $\hat{\epsilon}_i$ are our empirical residuals $\hat{\epsilon}_i \equiv Y_i - \hat{m}_{Z_i = j}(X_i)$.
Now we see that our variation is dictated primarily by the $\hat{\epsilon}_i$, which are more precise stand-ins for the true residuals in \eqref{eq:weighted_error}. We therefore construct variance estimates for the model-assisted estimator using the empirical residuals $\hat{\epsilon}_i$ in place of the outcome $Y_i$.

Following our assumption on the form of the quality surface \eqref{eq:quality_curve_linear} as $\hat{m}_j = \hat{\alpha}_j + \hat{\beta}_j \cdot \phi(x)$, we could estimate $\hat{m}_j$ via linear regression separately within each hospital. Unfortunately, this is difficult for those hospitals with relatively small surgery patient populations or those hospitals that tend to differ from the target population.

We therefore fit a model with hospital-specific intercepts but common coefficients across hospitals, $\hat{m}_j(x) = \hat{\alpha}_j + \hat{\beta} \cdot \phi(x)$ \citep{normand2016league}.
This allows the model to share information on the relationship between the covariates and the outcome, while still allowing for systematic differences in hospital quality.
More elaborate pooling procedures are possible, e.g., directly using the hierarchical Bayesian modeling indirect standardization approaches \citep{george2017mortality}.
Plugging the fixed effect model into Equation~\eqref{eq:bias_correct}:
\begin{equation}
  \label{eq:bias_correct_linear}
  \hat{\mu}_j = \sum_{Z_i = j}\hat{\gamma}_i Y_i + \underbrace{\hat{\beta} \cdot \left(\bar{\phi} - \sum_{Z_i = j} \hat{\gamma}_i \phi(X_i) \right)}_{\text{adjustment for remaining imbalance}}.
\end{equation}
The hospital fixed effects $\hat{\alpha}_j$ drop out due to the sum-to-one constraint, leaving us with only the pooled coefficients $\hat{\beta}$.

In general, bias adjustment is more aggressive for hospitals with larger imbalances. This is be especially true for hospitals with smaller effective sample sizes, as there are fewer patients available for trying to match the population distribution; for these hospitals, the bias adjustment leads to some extrapolation away from the observed patient mix in order to achieve better balance \citep{benmichael2019_ascm}. The adjustment are smaller for hospitals with excellent balance, e.g., hospitals with large patient populations or a high degree of overlap. Regardless of hospital type, the adjustment increases both precision and estimated precision. 

In sum, we can use an outcome model to incorporate additional risk adjustment into our estimates of hospital quality, albeit at the price of additional model dependence. We refer to risk adjustment using weights alone as \emph{weighted risk adjustment} and risk adjustment using both weights and additional outcome modeling as \emph{bias-corrected risk adjustment}.

\subsection{Partially Pooling Hospital-Specific Estimates}
\label{sec:bayes}

Thus far, our approach estimates hospital-specific means $\hat{\mu}_j$ in relative isolation. These estimates can be unstable, especially for the smaller hospitals and those hospitals with low effective sample size. Following standard practice in hospital quality research, we therefore partially pool the estimates via a hierarchical Bayesian model \citep{normand2007statistical,iezzoni2012risk,george2017mortality}. We can use this approach either with or without the bias correction step in Section \ref{sec:covariate_adjustment}.

The important change from the no pooled estimate is that we now assume that the hospital-specific complication rates are drawn from an underlying random effects distribution, $G$:
\begin{align*}
\hat{\mu}_j &\sim N(\mu_j, \widehat{\text{se}}_j^2 ) \\
\mu_j &\sim G,
\end{align*}
with estimated estimated standard error, $\widehat{\text{se}}_j = \hat{\sigma}_{\text{pool}} \Big/ \sqrt{ n^{\text{eff}}_{j} }$. 
This is a ``modular'' Bayesian procedure that treats $\widehat{\text{se}}_j$ as known, which avoids some complications that arise from estimating hospital-specific variances in a fully Bayesian setup  \citep{jacob2017better}. Nonetheless, it is straightforward to extend this approach to a fully Bayesian or empirical Bayes setup, such as triple goal estimation \citep{shen1998triple,paddock2006flexible}.

This approach differs from other generalized linear mixed models used for assessing hospital quality \citep{normand2007statistical, normand2016league}.
These approaches tend to partially pool parameters in the outcome model (e.g., varying intercept models) rather than on the mean itself.
We view our approach as more transparent, since the analyst controls the level of partial pooling directly.


\section{Simulation Study}

Next, we conducted a simulation study to explore the benefits of additional regression adjustment compared to weighting alone. For each scenario considered, we first generate a fixed set of $J$ hospitals, with each hospital having a specific distribution of patients, relationship of patient characteristics and outcomes, and overall quality score. We then repeatedly sample patients from each hospital, generate a binary adverse effect for those patients, and fit the resulting data using three strategies: simple averaging (no adjustment), our weighting adjustment, and our weighting adjustment with an additional covariate adjustment. We then compare the sets of resulting hospital specific quality estimates to the true quality scores and measure variability, bias, and root-mean squared prediction error (RMSPE). Overall we verify, as expected, that in contexts where hospitals serve different patient demographics, weighting does reduce bias, but this comes at a cost of increased variance. Regression adjustment provides additional benefit by giving a further bias reduction. The additional regression adjustment does not increase estimator variance as the covariates---all at the individual level---are estimated with high precision given the large sample size. We now provide specific details of our model and simulation, and then present the results.

\subsection{The Data Generation Process}

We characterize each hospital with three independent variables, $u_0, u_1, u_2 \stackrel{\text{iid}}{\sim} \text{Unif}[-0.5,0.5]$, representing latent hospital characteristics. We denote this set of variables as $u_k$. These variables jointly drive four main aspects of the data generation process: hospital size, hospital quality, hospital patient population served, and the hospital-specific outcome-covariate relationship of the hospital's patients. We have four primary simulation factors: $\bar{\alpha}$, $\sigma^2_{\alpha}$, $\bar{\beta}$, and $\sigma^2_{\beta}$. Conceptually, $\bar{\alpha}$ is the average hospital quality (i.e. is the mean hospital intercept in a linear model), controlling what proportion of the patients at a median hospital would have a complication, and $\sigma^2_{\alpha}$ is the variance of the individual hospital-level intercepts. The relationship between the covariates and the outcome is controlled by $\bar{\beta}$ and $\sigma^2_{\beta}$, with $\bar{\beta}$ controlling how strongly the patient level covariates predict patient level risk on average and $\sigma^2_{\beta}$ controlling the variation across hospitals in the strength of this relationship. A non-zero $\sigma^2_{\beta}$ allows different hospitals to have different relationships between their patient covariates and outcomes.

Within each hospital $j$, a patient $i$, has as a vector of 7 covariates, and the following patient-level risk $\pi_{ij}$:
\begin{equation}
	\logit( \pi_{ij} ) = \bar{\alpha} + \alpha_j  + \bar{\beta}v'(X_{ij} - \bar{X}) + (\beta_{j}-\bar{\beta})v'X_{ij}, \label{eq:patient_risk}
\end{equation}
where $v = (0.4, 0.3, 0.4, 0.2, 0.2, 0.2, 0.2)$ is a vector of coefficients for our 7 covariates, $X_{ij}$ is the demographic vector of covariates for patient $i$ in hospital $j$, and $\bar{X}$ is expected average of the $X_{ij}$ across the full population.
The $\alpha_j$ is the deviation of how much more (or less) effective hospital $j$ is than average, and $\beta_j$ controls how much stronger (or weaker) the covariate-outcome relationship is. We link the $u_k$ to $\alpha_j$ with 
$$\alpha_j = \bar{\alpha} + \sigma_{\alpha}4(u_0 + u_1 + u_2 - 1.5)$$
and
$$\beta_j = \bar{\beta} + \sigma_{\beta} 6(u_0 + u_1 - 1),$$
where the centering of $-1.5$ and $-1$ and scaling of $4$ and $6$ to make the $\alpha_j$ and $\beta_j$ centered at zero with variances of $\sigma^2_\alpha$ and $\sigma^2_\beta$. 

Given a set of $J$ hospitals, we generate a sample of $N = 80 J$ patients across the hospitals, with each hospital's size being proportional to $u_0 + 0.3$, allowing some hospitals to be up to $4 1/3$ times larger than others, with an average size of 80. For each patient, we generated covariates with varying relationships to the latent $u_k$, which induces different patient populations across the different hospitals. Our first covariate is a count, with $X_1 \sim \text{Pois}(\lambda)$, with $\lambda = 1 + 1.5\exp^{-1}(u_0)$, where $\exp^{-1}()$ is the quantile function of the exponential distribution.  Our second and third characteristics are binary indicators with $X_2 \sim Bern( \frac{1}{3}( u_1 + u_2 + 0.5 ) )$ and $X_2 \sim Bern( \frac{1}{3}( u_0 + u_1 + u_2 ) )$.
The four additional binary covariates are not related to the hospital characteristics and are all drawn as independent $Bern( 1/2 )$. We then calculate patient level risks using Equation~\ref{eq:patient_risk}, and finally patient outcomes $Y_{ij}$ as Bernoulli draws with the given $\pi_{ij}$. In sum, patient level covariates predict the risk of a complication; however, there is a hospital level component that varies from hospital to hospital in a stochastic fashion. 

\subsection{Simulation Implementation}

We seek to understand how well our proposed methods recover the true hospital quality given the randomness of the samples. As such, in our simulation, we use our model to generate an initial fixed population of hospitals and patients. In each simulation, we then take random samples with replacement from this population, keeping patient hospital membership fixed. This induces variability in the patients we use to estimate the hospital level outcomes. In each simulation, we apply our estimation approaches to estimate hospital quality for each hospital. We compare these values to those based on hospital level estimates without risk adjustment.

We focus on $\bar{\beta}$ as the primary factor of our simulation. As $\bar{\beta}$ gets larger, the bias in the na\"ive estimates of hospital quality should increase, because patient level risk will depend to a greater extent on the covariates, causing hospitals serving different types of patients to diverge.
In our simulations, we used ten different values for $\bar{\beta}$ that varied from 0 to 3 in increments of 2/3. We set $\bar{\alpha} = -1$ and vary $\sigma^2_{\alpha}$ and $\sigma^2_{\beta}$ across 0 and 1. For each set of simulation parameters, we generate a fixed set of hospitals and repeat the simulation 1,000 times, resampling the patients with each trial. We fix the hospitals in order to be able to directly estimate the RMSPEs for the individual hospital-specific quality estimates. For each simulation, we calculated the standard error, bias, and RMSPE for each hospital, and then average across hospitals.

\begin{figure}
  \centering
    \includegraphics[width=\textwidth]{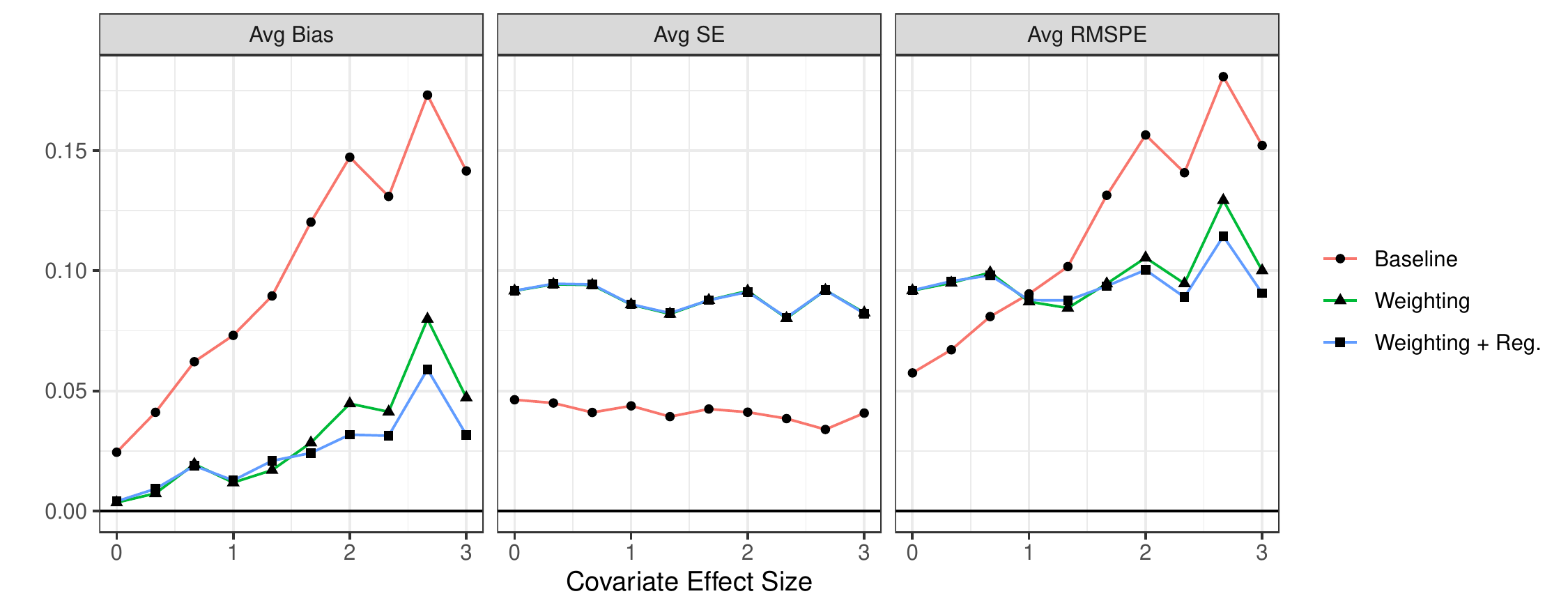}
    \caption{Average bias, standard error, and RMSPE across all hospitals and simulation runs. $X$-axis is $\bar{\beta}$. Baseline represents unadjusted hospital level results.}
  \label{fig:sim}
\end{figure}

\subsection{Results}

Figure~\ref{fig:sim} reports results for when $\sigma^2_{\alpha} = \sigma^2_{\beta} = 1$. See the appendix for results with the full set of simulation parameters. As expected, the unadjusted results always have higher bias than when we adjust via weighting (we still have bias when $\bar{\beta}=0$ due to the individual $\beta_j$ varying and being correlated with hospital size). Adjustment does not, in our circumstance, solve all our problems; the varying relationship between covariate and outcome means reweighting only reduces bias but does not completely eliminate it. That being said, most of the bias is eliminated across all values of $\bar{\beta}$.

Despite the bias reduction, weighting can have a higher RMSPE than no adjustment when the covariates are not generally predictive (i.e., $\bar{\beta}$ is low); in these cases the bias-variance trade-off of weighting is too expensive in terms of variance. We see the variance cost is essentially independent of the covariate-outcome relationship, as illustrated by the flat SE curves. A higher RMSPE may be desirable: without bias reduction, the subsequent step of estimating cross-hospital variation could be undermined as bias is not be taken into account in the variance decomposition in the shrinkage estimation stage. Overall, depending on the purpose of the hospital-specific estimates, reducing bias may be worth the increased variance, even if the overall RMSPE is higher.

The addition of regression adjustment further reduces bias as the strength of the average relationship between covariate and outcome increases. This reduction appears to be essentially free: the standard error plot shows no increase in variance of the regression+weighting vs. weighting only. As the regression adjustment is estimating parameters fully pooled across hospitals, we can leverage the full sample of individuals to estimate these coefficients with high precision. Given the bias reduction and no increase in variance, we see an overall reduction in RMSPE. For the higher $\bar{\beta}$, RMSPE reduction is a bit more than 10\%.


\section{Hospital Performance on Postoperative Complication in General Surgical Performance}
\label{sec:results}

\subsection{Setup}
\label{sec:results_setup}

We now apply our approach to estimate risk-adjusted complication rates for general surgery patients in Pennsylvania, New York, and Florida. 
Our first step is to build $\phi(x)$, the basis for assessing covariate imbalance; we set this to be a standardized version of the full set of covariates. Standardizing is important because dimensions of $\phi(x)$ with high variances will implicitly receive greater weight in the optimization problem. For binary covariates with estimated proportion $\hat{p} \geq 0.05$, we standardize by subtracting the mean and dividing by the standard deviation.  For binary variables with rare outcomes, $\hat{p} < 0.05$, we standardize by $\sqrt{0.05 \cdot 0.95}$ instead of $\sqrt{\hat{p}(1 - \hat{p})}$, which prevents extremely rare covariates from receiving too much weight in the optimization process. For continuous covariates, we standardize by subtracting the mean and dividing by the standard deviation. We also augment our $\phi(x)$ with one aggregate measure. In our context, many of our covariates are indicators for relatively rare comorbidities, and matching on all of them separately may be difficult. We therefore add a key summary measure of this risk, the number of total comorbidities for each patient, and include this as an additional covariate in $\phi(x)$. Our setup is general and, in principle, we could generate a richer basis.

To assess the performance of our weighting procedure, we focus on the increase in precision and reduction in bias. For precision, we calculate the implied effective sample size, $n_j^{\text{eff}}$. For bias, we calculate the improvement in (weighted) covariate imbalance by hospital. For each hospital, we calculate $\overline{\phi(X)}_h$, the unadjusted covariate means, and $\overline{\phi(X)}_{h,w}$, the weighted means, using the weights from our procedure.
Next, we regress $Y_i$ on $\phi(X_i)$ to obtain a vector of regression coefficients $\hat{\eta}$, which give us variable importance weights. Using these estimates, the initial bias is $\Delta_h = (\overline{\phi(X)}_h - \overline{\phi(X)})'\hat{\eta}$, and the final bias is $\Delta_{h,w} = (\overline{\phi(X)}_{h,w} - \overline{\phi(X)})'\hat{\eta}$, for each hospital $h$. The first quantity is the estimated bias due to baseline differences in case mix; the second is the estimated remaining bias due to case mix after weighting. Using these two quantities, we then calculate the Percent Bias Reduction (PBR):
\[ 
PBR = 100\% \times \left[ \frac{1}{H} \sum_h |\Delta_{h,w}| \; \Big/ \; \frac{1}{H} \sum_h |\Delta_{h}| \right]  
\]
\noindent This measure describes the change in bias due to risk standardization while also accounting for the strength of the association between the different covariates and the outcomes.

\subsection{The bias-variance tradeoff and the role of $\lambda$}
\label{sec:lambda}

An important tuning parameter in our approach is $\lambda$, the global hyper-parameter that controls the bias-variance tradeoff: when $\lambda$ is large the optimization problem prioritizes variance reduction and searches for more uniform weights; when $\lambda$ is small it instead prioritizes bias reduction, allowing extreme weights that can reduce $n_j^{\text{eff}}$ for each hospital. To investigate the role of $\lambda$, we estimated weights for each of a series of $\lambda$ values ranging between 0 and 3.5. For each $\lambda$ value, we computed the average PBR and average effective sample size across all hospitals. In this analysis, we focus on risk adjustment via weighting alone. We also set $u = 1$ (no constraint) which does not limit the amount of weight assigned to any one patient. In the supplemental materials, we present results for an analysis with $u = 0.2$ and found the results were unchanged.

Figure~\ref{fig:1} summarizes the results and demonstrates the trade-off between bias reduction and effective sample size. When $\lambda=0$ (no attempt to control variance) the estimated percent bias reduction is 80\% relative to the unadjusted estimate, with an average effective sample size less than 200. While this bias reduction is large, we cannot achieve perfect balance, especially for smaller hospitals. Conversely, when $\lambda=3.5$ bias reduction is approximately 47\% but the average effective sample size is nearly 1000, an increase of more than a factor of 3. The results in Figure~\ref{fig:1} suggest a value for $\lambda$ around 0.05, which decreases bias reduction from our maximum possible of 80\% by approximately three percentage points, but essentially doubles the average effective sample size.

As a comparison, we also implemented a template match. Following \citet{silber2014template}, we first created a template by taking 500 random samples from the patient population, each with a sample size of 300. Of these 500 random samples, we selected the sample with the smallest discrepancy between the random sample and the overall population means; this set of 300 patients serves as the template. Next, we matched patients from each hospital to the template, using optimal match with refined covariate balance. This approach is an extension of fine or near-fine balance designed to balance the joint distribution of many nominal covariates \citep{Pimentel:2015a}. In the match, we employed both a propensity score caliper and optimal subsetting.
For each hospital, we optimally match individual patients to patients in the template, dropping template patients that have no match within a given caliper. For each match, therefore, we can obtain up to 300 matched pairs, the size of the template; the number of matched pairs may be smaller, however, if only a subset of patients are comparable. The patients selected from each hospital serve as the risk adjusted population for that hospital, and the subsequent risk adjusted measure of complications is the proportion of complications in that matched sample. For the template match, we also calculate the PBR and average sample size across hospitals. We did not discard hospitals where the matches were poor as was implemented in \citet{silber2014hospital}. We did this to measure the performance of both methods across all hospitals in the sample.  

Figure~\ref{fig:1} shows that template matching, which does not directly minimize imbalance, does not have comparable performance: bias reduction is under 50\%, less than some of the most regularized $\lambda$ considered, and the average effective sample size is less than 300, only slightly above the fully unregularized $\lambda=0$. See the supplemental materials for more detailed results from this analysis.

\begin{figure}[htbp]
\centering
  \centering
  \includegraphics[width=0.5\textwidth]{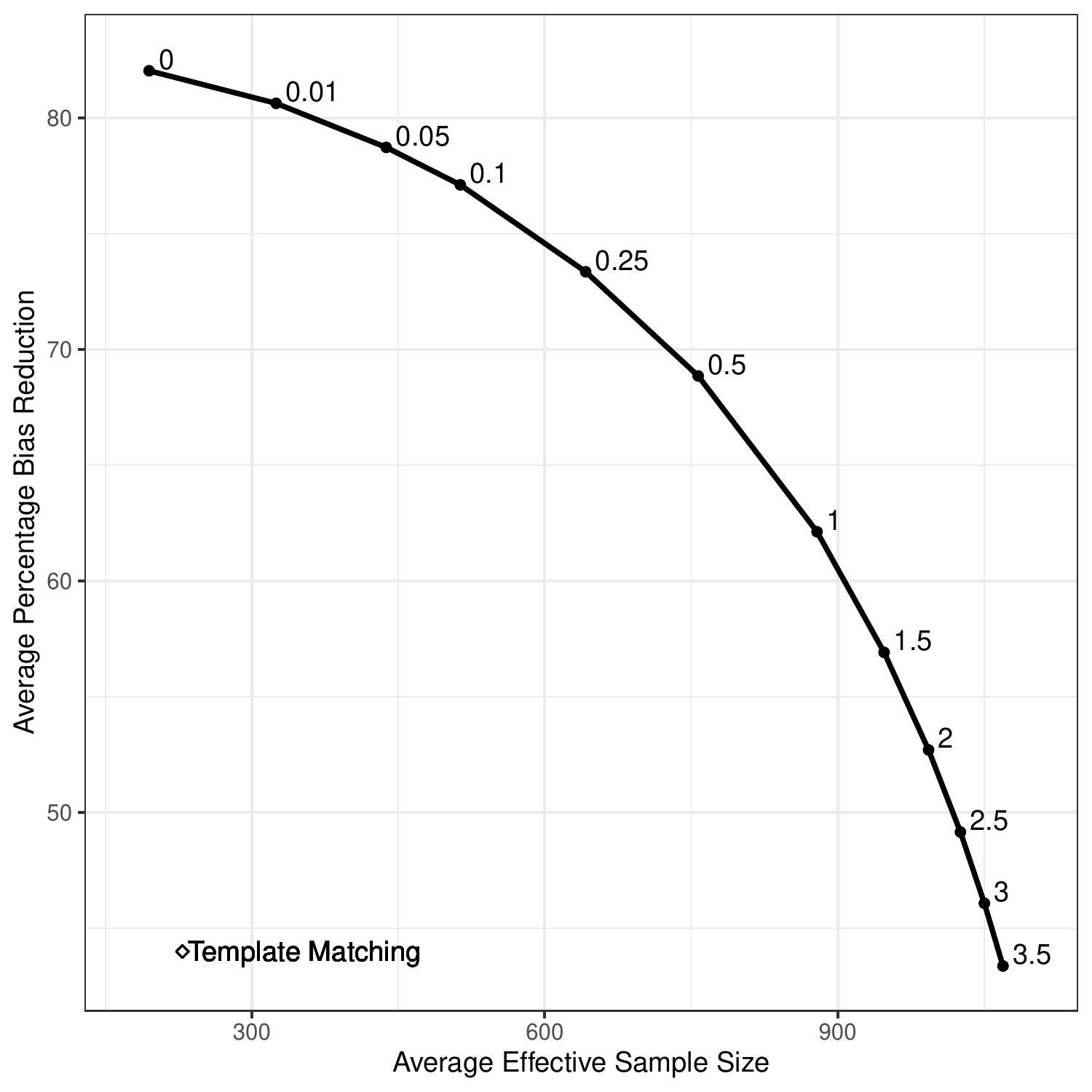}
  \caption{Estimated Bias-Variance Tradeoff as a Function of $\lambda$ values. Each dot represents the estimated Percent Bias Reduction and average effective size for template matching and for approximate balancing weights with different values of $\lambda$. The comparable values for template matching are shown in the bottom left, suggesting large gains in both bias reduction and effective sample size from using the proposed weighting approach relative to template matching.}
\label{fig:2}
\end{figure}

\begin{figure}
  \centering
  \begin{subfigure}[t]{0.48\textwidth}
    \includegraphics[width=\textwidth]{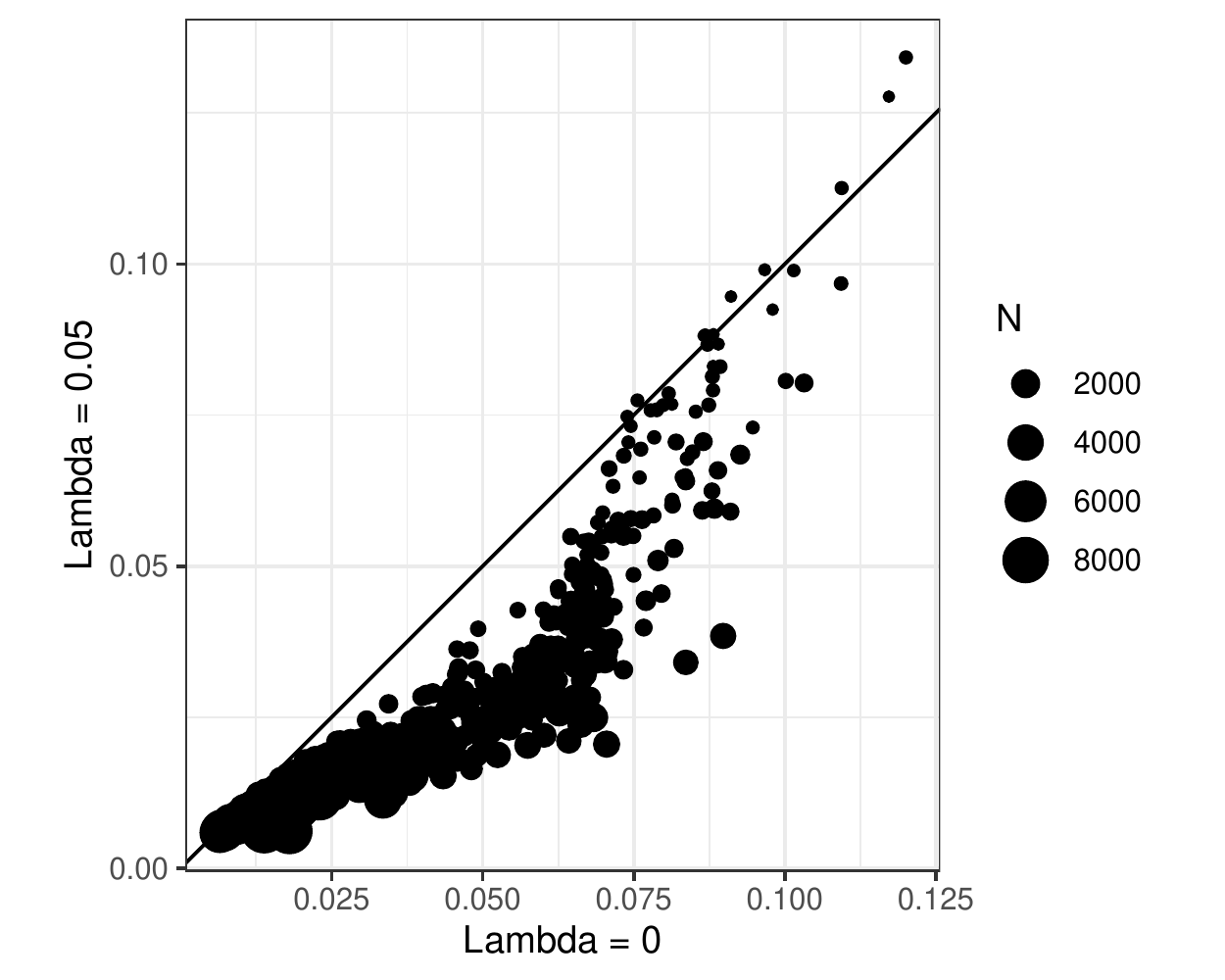}
    \caption{Standard Errors}
    \label{fig:se}
  \end{subfigure}
  \begin{subfigure}[t]{0.48\textwidth}
    \includegraphics[width=\textwidth]{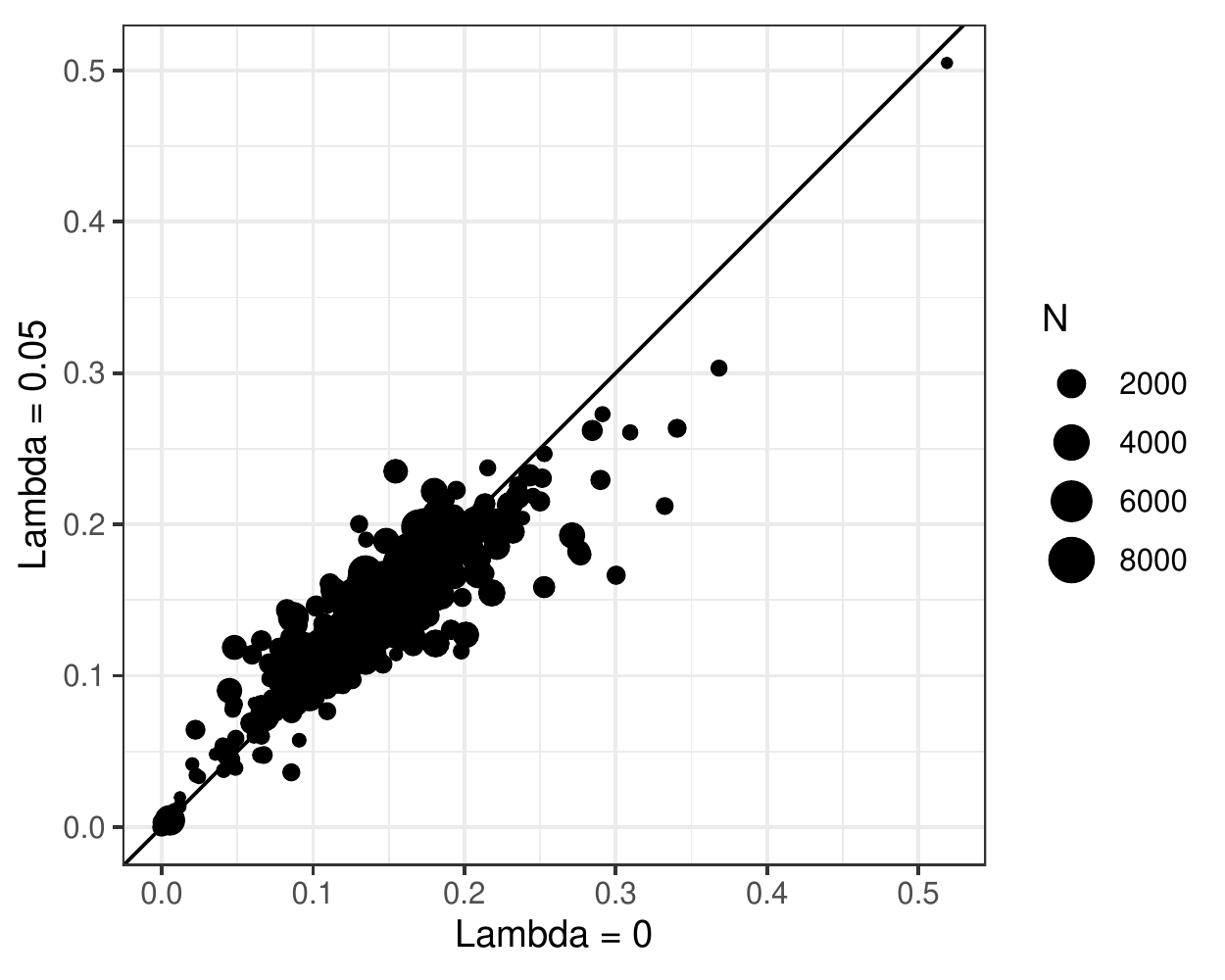}
    \caption{Risk Adjusted Estimates}
    \label{fig:pe}
  \end{subfigure}
  \caption{Scatterplots of hospital level standard errors and point estimates for $\lambda = 0.05$ and $\lambda = 0$.}
  \label{fig:lcomp}
\end{figure}

We next compared the estimated outcomes and estimated standard errors for $\lambda=0$ and $\lambda=0.05$. Figure~\ref{fig:lcomp} contains two scatterplots.
Figure~\ref{fig:se} plots the pairs of hospital standard errors under each evaluation. Most points are well below the 45 degree line, showing that the estimated standard errors are generally smaller when $\lambda = 0.05$. Figure~\ref{fig:pe} plots the risk adjusted estimated complication rates for each hospital. We observe that the estimates for both $\lambda = 0$ and $\lambda = 0.05$ agree for many hospitals, and agree on average, but for a few hospitals the differences are larger.

We then estimated the average bias reduction and average effective sample size for the two scenarios. When $\lambda = 0$, the bias reduction was 82\% with an average effective sample size of 194. When $\lambda = 0.05$, the bias reduction was 78.7\% with an average effective sample size of 438.
Overall, these results suggest that there are gains from selecting a value for $\lambda$ larger than 0: with a small increase in $\lambda$ we lost little in terms of bias reduction while more than doubling the effective sample size.

The results from this analysis raise the question of how users should select a value for $\lambda$ in applied clinical research. While we do not yet have a data-driven approach for selecting $\lambda$, the approach we use here seems reasonable for assessing the bias-variance tradeoff. More specifically, users can start with $\lambda = 0$ as a reference point, since this will maximally reduce bias. Users can then estimate additional fits with larger $\lambda$ values to find the point where bias reductions does not suffer but effective sample size is maximized. Importantly, such choices can be done without respect to outcome information, and therefore in a principled way.

\begin{figure}[hb]
\centering
\includegraphics[width=0.7\textwidth]{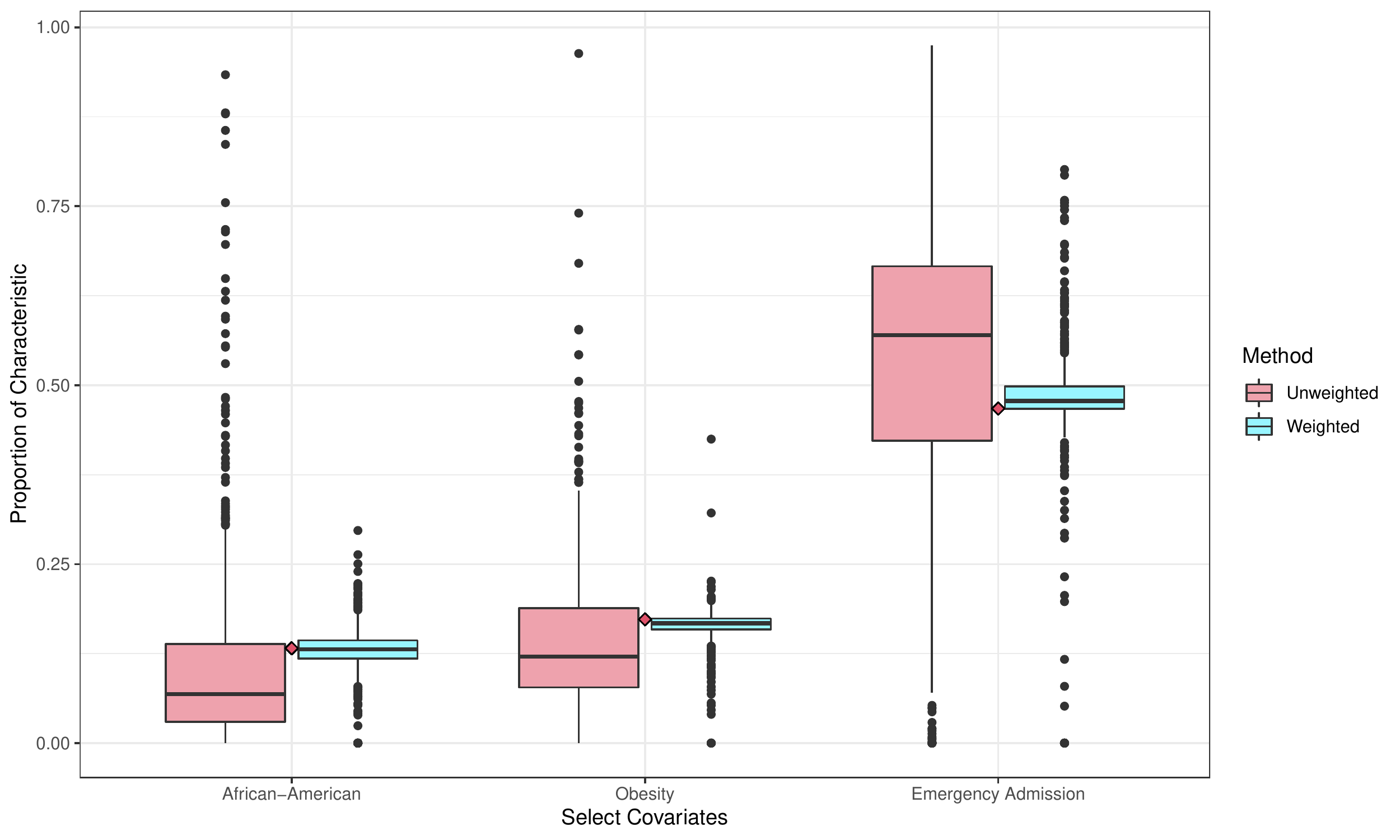}
\caption{Boxplots of Hospital Casemix Distributions Before and After Weighting. Diamond represents population mean for that covariate. $\lambda = 0.05$.}
\label{fig:box2}
\end{figure}

Finally, we demonstrate how the weights affect the distribution of covariates at the hospital level. Figure~\ref{fig:box2} shows box plots of hospital means before and after weighting (when $\lambda = 0.05$) for the three key covariates shown in Figure~\ref{fig:1}.  After weighting, the distributions of hospital means are clearly much closer to the population means, though some variation remains. 

\subsection{The role of risk adjustment on hospital quality}

\begin{figure}[htbp]
\centering
  \centering
  \includegraphics[width=.8\textwidth]{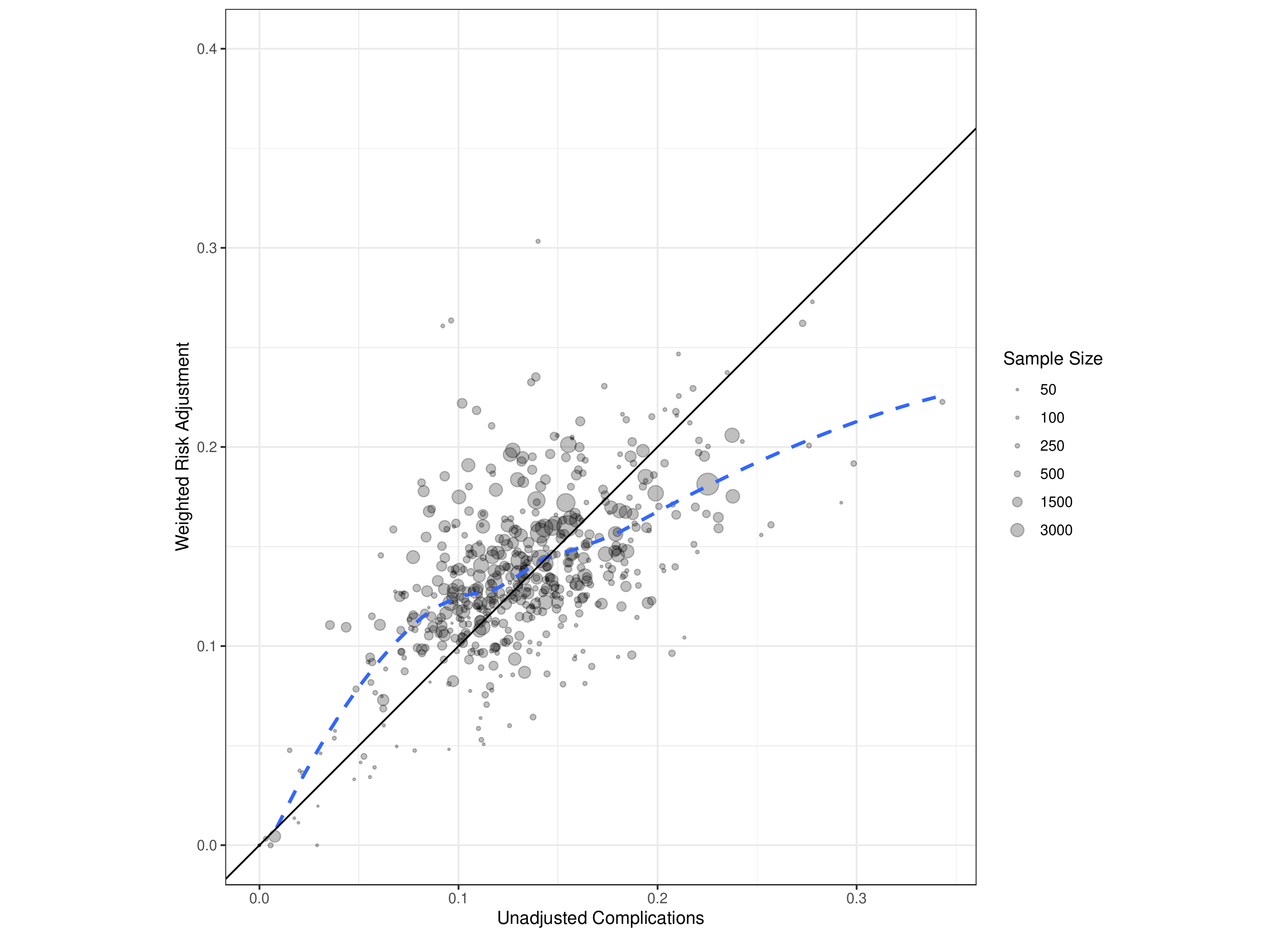}
  \caption{Scatterplot of the proportion of complications before risk adjustment against estimates after weighted risk adjustment. Dashed line is a loess fit.}
\label{fig:wgt.scatter}
\end{figure}

We now focus more directly on how risk adjustment alters hospital level outcomes. To that end, we examine the effects of weighted risk adjustment and then explore the utility of the additional bias-correction step. Figure~\ref{fig:wgt.scatter} shows the proportion of complications for each hospital before risk adjustment against the proportion of complications after weighted risk adjustment. Hospitals close to the 45 degree line saw little change; hospitals off the line were changed more. Generally, weighted risk adjustment induces relatively small changes, although several smaller hospitals changed quite a bit. The nonparametric trend line shows that in the middle of the distribution, risk adjustment tends to move complication rates up. We should also note that Figure~\ref{fig:wgt.scatter} reflects both actual variation as well as measurement error; we focus on accounting for measurement error with the shrinkage estimator below.

\begin{figure}[thbp]
\centering
  \centering
  \includegraphics[width=0.8\textwidth]{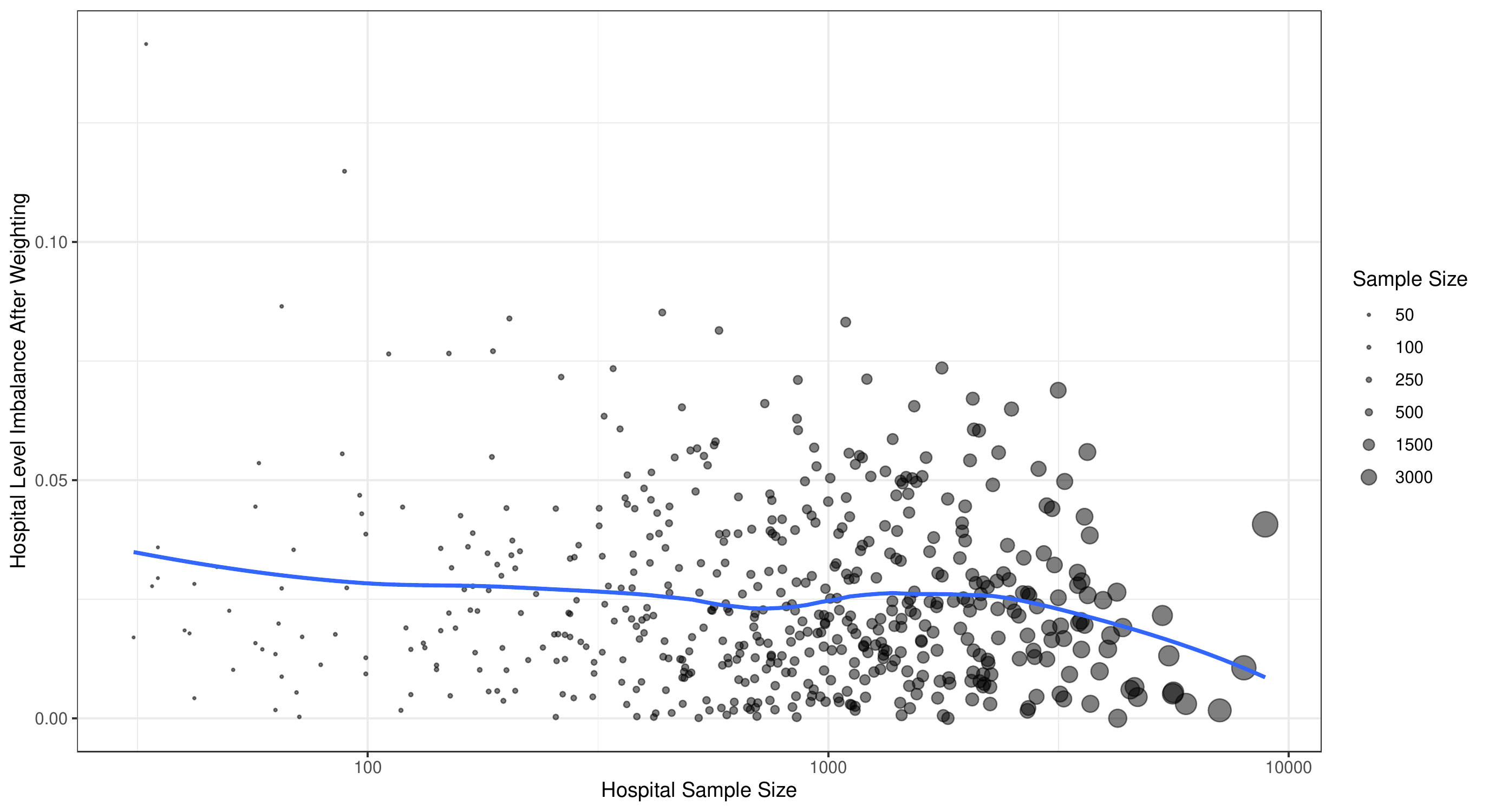}
  \caption{Scatterplot of residual hospital level bias and the logarithm of hospital sample size for $\lambda = 0.05$. Solid line is a loess fit.}
\label{fig:mod}
\end{figure}

The optimization is designed to re-weight hospitals to be as close to the target patient mix as possible, while maintaining some control over the dispersion of the weights.  Thus, depending on the original mix in a given hospital, it may not be possible to fully achieve exact balance on all the covariates. If imbalances remain, we can adjust for them using an outcome model.  Here, we explore the relationship between remaining imbalance and hospital sample size. Figure~\ref{fig:mod} shows the relationship between imbalance, as measured by $|\Delta_h - \Delta_{h,w}|$, the absolute difference between initial and final bias and hospital sample size. Overall, Figure~\ref{fig:mod} shows that some of the smallest hospitals have the largest residual biases after weighting. Thus we expect that these estimates will be more affected by model adjustment.

\begin{figure}[htbp]
\centering
  \centering
  \includegraphics[width=.85\textwidth]{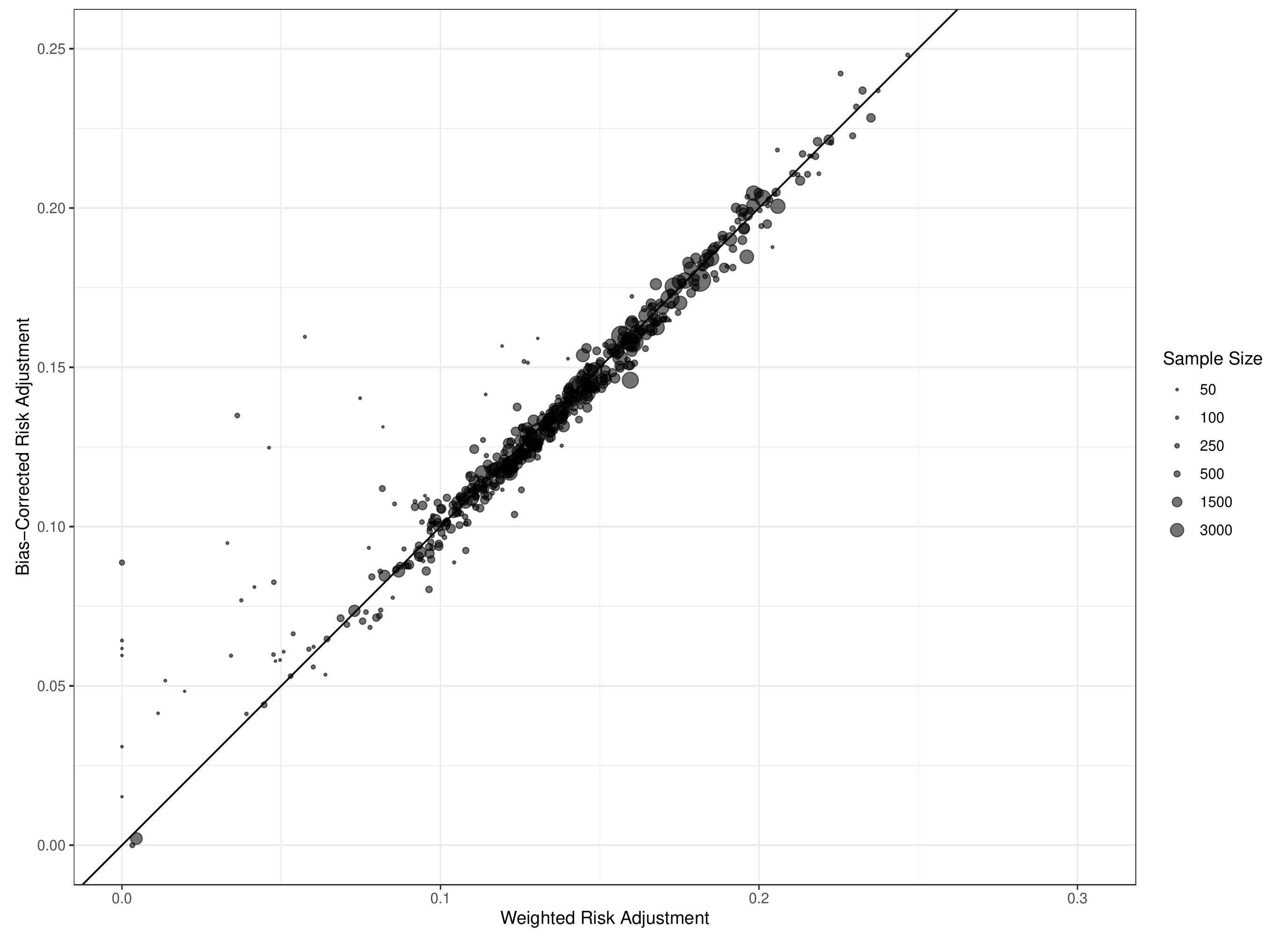}
  \caption{Scatterplot of the proportion of complications before risk adjustment against estimates that are weighted and adjusted.}
\label{fig:adj.scatter}
\end{figure}

We next plot the change in our hospital estimates due to bias correction. Figure~\ref{fig:adj.scatter} shows that estimates only change slightly for most hospitals, but that some smaller hospitals have substantial adjustment, as indicated by being far above the 45 degree line.  In particular, many small hospitals with very low estimated risk adjustments have substantial upward adjustments: the small sample size makes it difficult to re-weight existing patients to match the population. Critically, while bias correction can give a more accurate assessment of hospital quality, the adjustment is essentially an extrapolation based on the model.

Adjustment with our model should also improve the precision of our estimates, based on the model's predictive power. We calculated an $R^2$ value for our model by comparing the pooled variance of the residuals (weighted by our risk adjustment weights) to the pooled variance of the weighted outcomes: $1 - \hat{\sigma}^2_{\text{weighted pool}} \big/ \hat{\sigma}^2_{\text{bias corrected pool}} \approx 0.25$. Model adjustment, on average, removed 25\% of the variation within hospital. This variance reduction will lead to more precise standard errors due to removing variation that we can predict by specific case characteristics.

Finally, we investigate two hospitals where risk standardization resulted in large changes in the estimated complication rates. Hospital A had more than 500 general surgery patients,\footnote{We cannot disclose the specific hospital size in print per our data use agreement.} with an unadjusted complication percentage of 6.7\%, well below the sample average of 13.5\%. After we apply the weights, the risk-adjusted complication percentage increases to 16\%, above the sample average. Looking at this hospital's case mix makes this shift clear. The average age of the general surgery patients in this hospital at 51 is somewhat low compared to the overall population of 55, and these patients typically have only a single pre-existing medical condition. Moreover, only 15\% of the patients at this hospital are admitted for urgent procedures.  Hospital A confirms the value of risk adjustment: due to a relatively healthy patient population, hospital A appears to have a low complication percentage; once we risk adjust to match the population as a whole, the estimated complication percentage is much higher. 

Compare that to hospital B, which had more than 500 general surgery patients. The unadjusted complication percentage is 20.7\% and the risk adjusted complication percentage is 9.6\% ---  approximately a third lower than the percentage at hospital A. However, the patient case mix at hospital B is much different than for hospital A. The average age of the patient at this hospital is higher at 63 and patients typically have around three pre-existing medical conditions. Moreover, 60\% of the procedures at hospital B were urgent admissions. Thus, once we risk adjust to match the characteristics of the population as a whole, we find the estimated performance for hospital B actually improves. 

\subsection{Assessing the range of hospital quality}

We next assess the variation in estimated hospital quality, and how that variation changes when we adjust hospital complication rates to account for different patient mixes. To do this we use the ``Q statistic'' approach from meta-analysis \citep[see, e.g.,][]{hedges2001power}. The Q statistic is calculated as
\[ Q = \sum_{j=1}^n \frac{\left(\hat{\mu}_j - \bar{\mu}\right)^2}{\widehat{\text{se}}_j^2 + \tau^2 } , \]
where $\bar{\mu}$ is an estimate of the overall average outcome across hospitals (we use the simple mean) and $\tau^2$ is a hypothesized degree of cross-hospital variation in the true quality measures $\mu_j$.\footnote{To build intuition for this estimator, notice that we can decompose the difference of $\hat{\mu}_j$ and $\bar{\mu}$, as $\hat{\mu}_j - \bar{\mu} = (\hat{\mu}_j - \mu_j) + (\mu_j - \bar{\mu})$. The two terms in the denominator correspond to uncertainty in $\hat{\mu}_j - \mu_j$, captured by the estimation uncertainty $\widehat{\text{se}}_j$, and to uncertainty in $\mu_j - \bar{\mu}$, captured by the structural variation in hospital quality, $\tau$. 
For implementation, see the \texttt{blkvar} package: \url{https://github.com/lmiratrix/blkvar/}. }

Under the null hypothesis, $H_0: \tau = \tau_0$, the $Q$ statistic has an approximate $\chi^2_{n-1}$ distribution. We can then estimate $\tau$ using a Hodges-Lehman point estimate (corresponding to the $\tau$ with the largest $p$-value; here, the value where $Q = n-1$), and generate a confidence interval via test inversion.

We used this approach on three sets of hospital estimates: the raw mean outcomes of the hospitals without any adjustment, the mean outcomes of the hospitals after weighted risk adjustment, and the mean outcomes of the hospitals after both weighting and bias correction.
Results of these three analyses are summarized in Table~\ref{tab:q_table}. When we do not adjust for patient characteristics, the estimated standard deviation is over 5 percentage points. Under a Normality assumption, this suggests the complication rate for the middle 80\% of hospitals ranges from approximately 7\% to 20\%. 

When we standardize using weighting alone, the standard deviation falls sharply to 2.7pp. Relative to the raw estimates, this suggests that hospitals would have more similar outcomes if treating similar patients. We also see that 70\% of the variation in hospital quality is explained by the mix of patients, as measured by an $R^2$ type statistic of $R^2 = 1 - \sigma_{adj}^2 / \sigma_{raw}^2$.\footnote{This is a distinct quantity from how predictive individual covariates are for the outcome, which is the $R^2$ reported in the model adjustment section above.} Finally, the estimates are largely unchanged when we also incorporate bias correction. 

\begin{table}[bt]
  \begin{tabular}{l|rrcc}
  Estimates  & Grand Average & Std. Dev. & CI & 80\% Prediction Interval \\
    \hline
  Raw   				& 13.2\% 		& 5.1\% & (4.9\% -- 5.4\%) & 7\% -- 20\% \\
  Weighted Risk Adjustment 			& 13.5\% 		& 2.7\% & (2.4\% -- 2.9\%) & 10\% -- 17\% \\
  Bias-corrected Risk Adjustment & 13.6\% 		& 2.8\% & (2.6\% -- 3.0\%) & 10\% -- 17\%
  \end{tabular}
  \caption{Estimated average and standard deviation of hospital complication rates. First column is overall average across hospitals. The Std. Dev. is the estimated amount of variation in true complication rates across hospitals. The CI is the confidence interval for this amount of variation. The prediction interval estimates the range of the inner 80\% of hospitals, assuming normality in the complication rates. For the rows, Raw includes variation induced by different patient mix.}
  \label{tab:q_table}
\end{table}

\subsection{Results After Partial Pooling}

As described in Section~\ref{sec:bayes}, we now use a Bayesian hierarchical model to partially pool the hospital-specific estimates; we estimate this model using Stan, a Bayesian software package \citep{carpenter2017stan}.
We set the random effect $G$ as a simple Normal, $G = N(\alpha_\mu, \tau^2_\mu)$, consistent with prior work on hospital quality \citep{normand2007statistical, normand2016league}. Possible alternative parameterizations include a $t_7$ and a mixture of Normal distributions; see \citet{miratrix2020multisite}.
For Normal $G$ we impose a uniform prior over the random effect standard deviation, $\tau_\mu \in [0, \infty)$, and a uniform prior over the random effect mean, which we constrain to be in the unit interval, $\alpha_\mu \sim \text{Unif}[0,1]$, since we focus on binary outcomes. Results are largely unchanged with other prior choices.

\begin{figure}
  \centering
  \begin{subfigure}[t]{0.45\textwidth}
    \includegraphics[width=\textwidth]{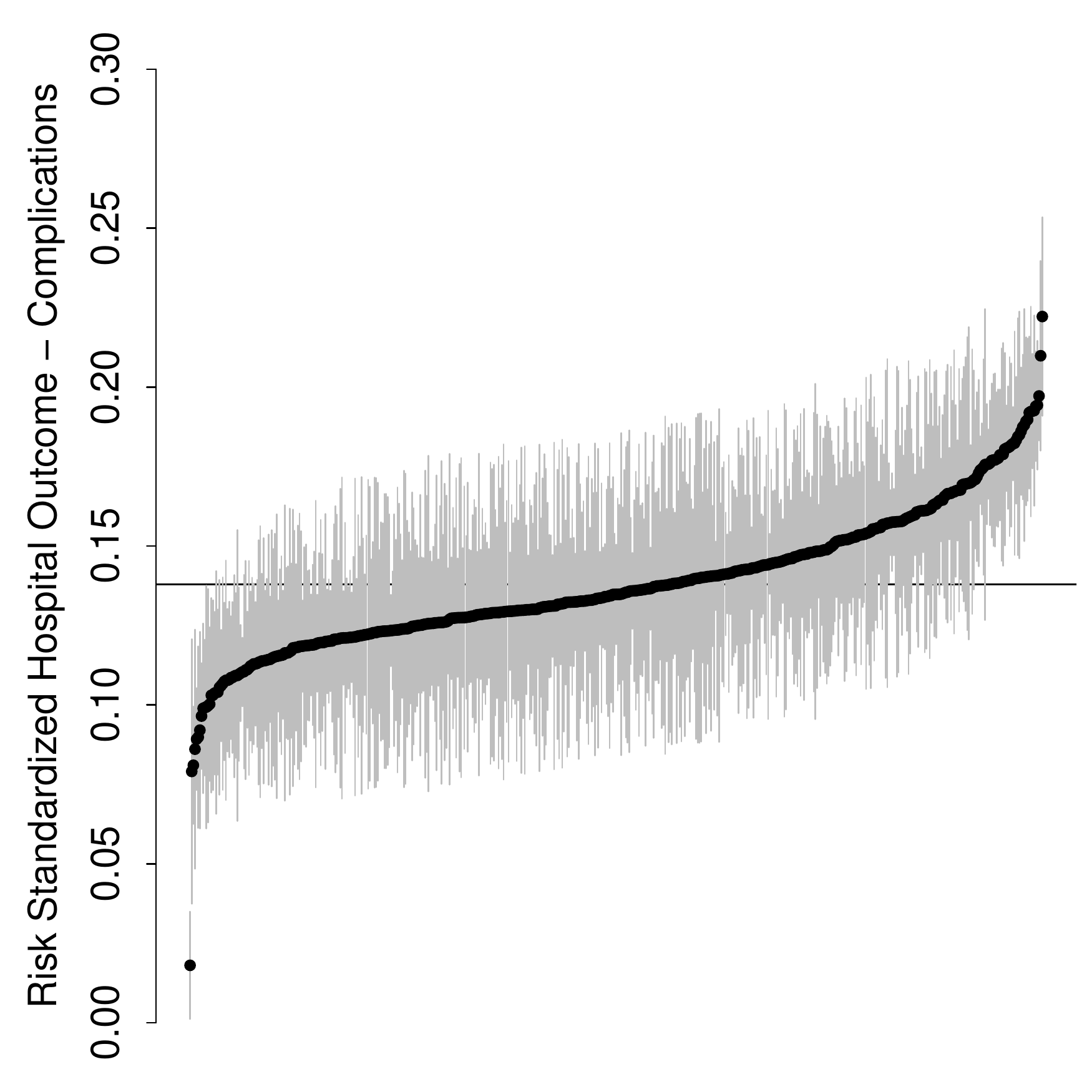}
    \caption{Catepillar Plot of Risk Standardized Complications for $\lambda = 0.05$}
    \label{fig:cat}
  \end{subfigure}
  \begin{subfigure}[t]{0.45\textwidth}
    \includegraphics[width=\textwidth]{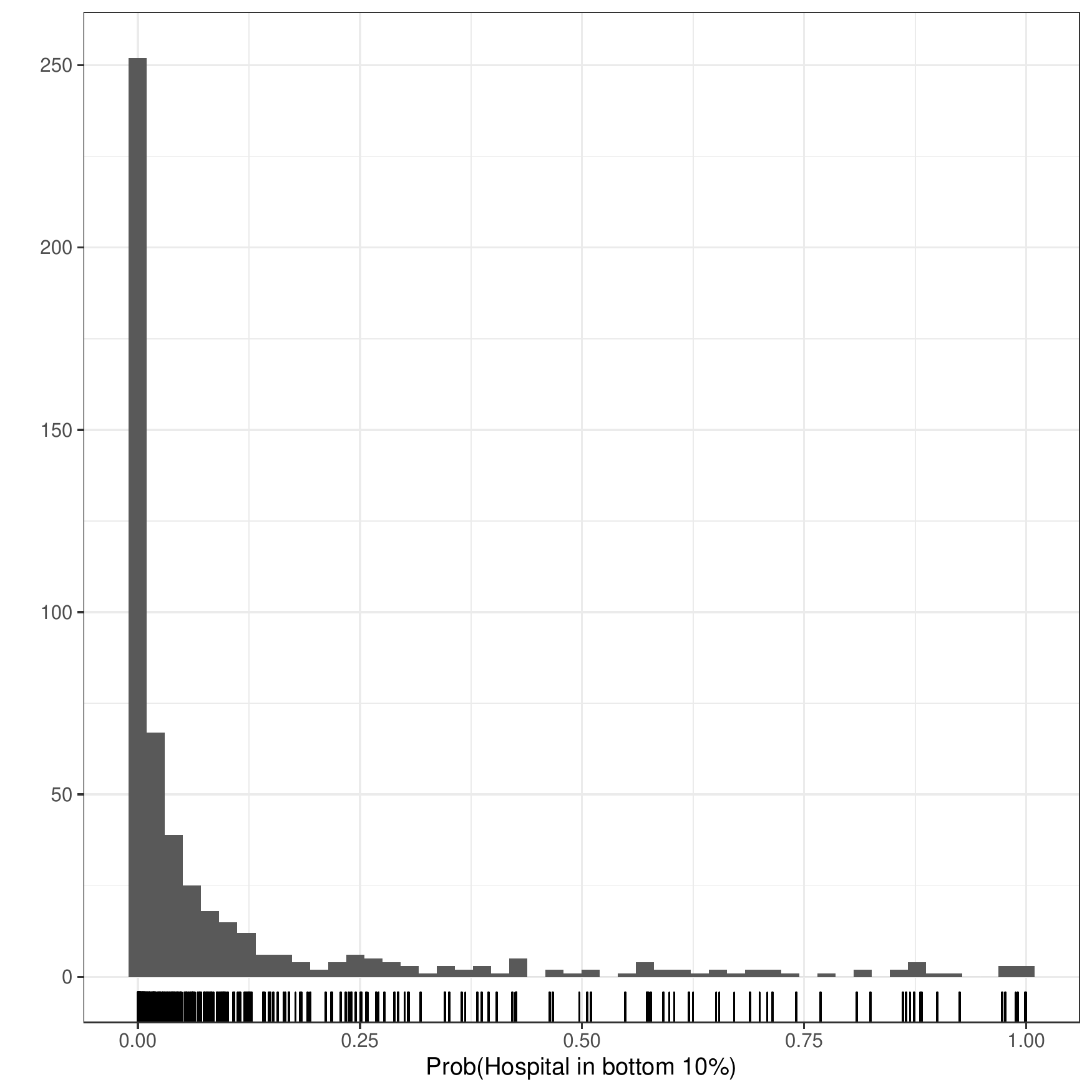}
    \caption{Histogram of the probability of being in the bottom 10\% of ranks $\lambda = 0.05$}
    \label{fig:prob}
  \end{subfigure}
  \caption{Hospital quality results after applying Bayesian shrinkage.}
  \label{fig:bayes}
\end{figure}

Figure~\ref{fig:cat} shows the posterior means and corresponding 95\% uncertainty intervals for the set of $\mu_j$, the risk-standardized hospital complication rates.\footnote{See \citet{paddock2006flexible} for a discussion of alternative approaches to summarizing the posterior in terms of the ``triple goals'' of estimating hospital-specific means, hospital-specific ranks, and the overall distributions.} 
We see variation in both the point estimates as well as the width of the hospital-specific uncertainty intervals. While there is a large mass of hospitals in the center of the distribution, there are clearly some hospitals with consistently above- or below-average estimated complication rates.

While the primary aim of our analysis is to produce risk standardized measure of hospital performance, risk adjustment is also used to identify institutions that are outliers \citep{ridgeway2009doubly}. For example, hospitals that are identified as underperforming may be targeted for quality improvement efforts. The Bayesian hierarchical model we fit can also be used for this purpose. Figure~\ref{fig:prob} shows the posterior probability that each hospital is in the highest decile --- that is, the \emph{worst} performing 10 percent --- of (standardized) surgical complication rates. For the vast majority of hospitals, the probability of being in this ``danger zone'' is quite low: 98.5\% of hospitals have a less than 10\% chance of being in this low performing group.

Some hospitals, however, are very likely to be low-performing: there are 9 hospitals that have at least a 90\% chance of being in this low performing group, 4 of which with at least a 99\% chance. Among the hospitals with at least a 90\% chance of being a low performer, the average adjusted complication rate was 23\% --- relative to 13.5\% overall.  Moreover, the average patient volume in this group of hospitals was over 2,700 patients, suggesting that, at least in our data, the low performing hospitals are not the lowest volume hospitals. 

\section{Conclusion}

Methods of risk adjustment are widely used to compare the performance of hospitals and physicians. Here, we develop a new method of direct standardization based on weighting. We treat each hospital as a sample from the overall patient population and find weights such that the re-weighted hospital patient mix matches the overall population. We obtain these weights via a convex optimization problem that trades off covariate balance and effective sample size. Finally, we applied our approach to data on general surgery in Pennsylvania, Florida, and New York.

This approach to risk adjustment offers several critical advantages. The risk adjusted outputs are readily interpretable. Principled methods of variance estimation are easily adapted from the literature on survey sampling and weighted regression. Compared to other direct standardization approaches, risk adjustment via weighting substantially reduces bias. We also found large increases in effective sample size for a slight increase in possible bias. We proposed a bias correction approach to incorporate outcome modeling as well. Finally, our method of direct standardization can also be combined with shrinkage methods to account for the variation in hospital size when comparing hospitals to each other and identifying high and low performing hospitals. Overall, the estimation process is not computationally intensive, and requires little user input outside of selecting the penalty. Estimating a set of weights for over 600,000 patients required less than five minutes on a desktop computer. Template matching, by contrast, required fine tuning of over five hundred different matches, and was a much more time consuming process.

We can extend the proposed approach to allow for a richer covariate basis, including interactions and higher-order terms, as well as to prioritize balance in some covariates (see Section \ref{sec:results_setup}). Another strategy is to use external data to fit an outcome model and then use our approach to balance the predicted value from that model \citep{damour2019reducer}. We expect that this will be a fruitful direction for future work. Currently we only use the weights to target population level means. We could easily apply this procedure to target a specific subset of patients, such as the average African-American patient in the population. Alternatively, the target for balance need not be the overall patient population. We could, instead, target the patient mix of a specific hospital. Parameter selection could also be optimized to find an optimal tradeoff between bias reduction and effective sample size. We could also explore best practice in terms of the application of the shrinkage methods.  



\clearpage
\renewcommand{\refname}{Bibliography \& References Cited}

\singlespacing
\bibliography{hosp_stand_bib}

\end{document}